\documentclass[aps,twocolumn,pra,superscriptaddress,showpacs,tightenlines]{revtex4-1}
\usepackage{amsmath}
\usepackage{amsfonts}
\usepackage{color}
\usepackage{epsfig}
\usepackage{graphicx}
\usepackage{txfonts}
\usepackage[colorlinks,citecolor=blue]{hyperref}
\begin{document}
\title{Thermal-noise-resistant optomechanical entanglement via general dark-mode control}

\author{Jian Huang}
\affiliation{Key Laboratory of Low-Dimensional Quantum Structures and Quantum Control of Ministry of Education, Key Laboratory for Matter Microstructure and Function of Hunan Province, Department of Physics and Synergetic Innovation Center for Quantum Effects and Applications, Hunan Normal University, Changsha 410081, China}

\author{Deng-Gao Lai}
\email{Corresponding author: denggaolai@foxmail.com}
\affiliation{Theoretical Quantum Physics Laboratory, RIKEN Cluster for Pioneering Research, Wako-shi, Saitama 351-0198, Japan}
\affiliation{Key Laboratory of Low-Dimensional Quantum Structures and Quantum Control of Ministry of Education, Key Laboratory for Matter Microstructure and Function of Hunan Province, Department of Physics and Synergetic Innovation Center for Quantum Effects and Applications, Hunan Normal University, Changsha 410081, China}

\author{Jie-Qiao Liao}
\email{Corresponding author: jqliao@hunnu.edu.cn}
\affiliation{Key Laboratory of Low-Dimensional Quantum Structures and Quantum Control of Ministry of Education, Key Laboratory for Matter Microstructure and Function of Hunan Province, Department of Physics and Synergetic Innovation Center for Quantum Effects and Applications, Hunan Normal University, Changsha 410081, China}

\begin{abstract}
Quantum entanglement not only plays an important role in the study of the fundamentals of quantum theory, but also is considered as a crucial resource in quantum information science. The generation of macroscopic entanglement involving multiple optical and mechanical modes is a desired task in cavity optomechanics.  However, the dark-mode effect is a critical obstacle against the generation of quantum entanglement in multimode optomechanical systems consisting of multiple degenerate or near-degenerate mechanical modes coupled to a common cavity mode. Here we propose an auxiliary-cavity-mode method to enhance optomechanical entanglement in a multimode optomechanical system by breaking the dark-mode effect. We find that the introduction of the auxiliary cavity mode not only assists the entanglement creation between the cavity mode and the mechanical modes, but also improves the immunity of the optomechanical entanglement to the thermal excitations by about three orders of magnitude. We also study the optomechanical entanglement in the network-coupled optomechanical system consisting of two mechanical modes and two cavity modes. By analyzing the correspondence between the optomechanical entanglement and the dark-mode effect, we find that optomechanical entanglement can be largely enhanced once the dark mode is broken. In addition, we study the mechanical entanglement and find that it is negligibly small. We also present some discussions on the experimental implementation with a microwave optomechanical setup, on the relationship between the dark-mode-breaking mechanism and the center-of-mass and relative coordinates,  and on the explanation of the important role of the dark-mode breaking in the enhancement of optomechanical entanglement. Our results pave the way towards the preparation of entangled optomechanical networks and noise-resistant quantum resources.
\end{abstract}

\maketitle

\section{Introduction\label{introduce}}

As one of the cornerstones of quantum theory, quantum entanglement is a physical phenomenon that the measurement of one part of a quantum system can determine the state of the other part, no matter how far the two parts are separated~\cite{Horodecki2009}. Up to now, quantum entanglement has been widely used in various quantum technologies, such as quantum computation, quantum communication, quantum sensing, and quantum information processing~\cite{Nielsen2000,Duarte2021}. In particular, quantum entanglement has been observed in numerous physical systems, including photons~\cite{Pan2012}, trapped ions~\cite{Leibfried2003}, atoms~\cite{Raimond2001}, and superconducting qubits~\cite{Xiang2013,Gu2017}, ranging from microscopic-scale objects to macroscopic integrated devices.

In parallel, owing to the great progress in the ground-state cooling of mechanical oscillators~\cite{Wilson-Rae2007,Marquardt2007,Chan2011,Teufel2011a} and the realization of strong linearized optomechanical coupling~\cite{Groblacher2009,Teufel2011b,Verhagen2012}, the cavity optomechanical system has become a powerful platform for the study of fundamental quantum physics and the manipulation of macroscopic quantum objects~\cite{Kippenberg2008,Aspelmeyer2014,Metcalfe2014}. Recently, some schemes for generation of quantum entanglement via optomechanical interfaces have been proposed ~\cite{Mancini2002,Vitali2007prl,Vitali2007jpa,Paternostro2007,Genes2008,Hartmann2008,Genes2011,Jiao2020}.
Moreover, macroscopic mechanical entanglement between two mechanical oscillators has been experimentally demonstrated~\cite{Riedinger2018,Ockeloen-Korppi2018,Kotler2021,Lepinay2021}. Meanwhile, various schemes have been proposed for improving quantum entanglement in cavity optomechanical systems, such as conditional measurements on optical modes~\cite{Pirandola2006,Borkje2011,Abdi2012,Flayac2014}, reservoir engineering~\cite{Tian2013,Wang2013}, coherent feedback~\cite{Li2015,Li2017}, and the transfer of entanglement from optical modes to mechanical modes~\cite{Zhang2003,Ge2013}.

Recently, multimode optomechanical systems involving two or multiple mechanical modes~\cite{Riedinger2018,Ockeloen-Korppi2018,Kotler2021,Lepinay2021,Bhattacharya2008,Massel2011,Stannigel2012,Xuereb2012,Huang2013,Xu2013,Liao2014,Spethmann2016,Piergentili2018,Lai2018,Yang2020,Lai2021a,Lai2021b} have become a new research hotspot. This is because many interesting physics have been found in these systems, such as quantum synchronization~\cite{Mari2013,Zhang2015} and quantum many-body effects~\cite{Heinrich2011,Ludwig2013,Xuereb2014}.
Motivated by recent interest in multimode optomechanics, the generation of quantum entanglement in multiple-mechanical-mode optomechanical systems becomes an interesting task. However, it remains a great challenge to generate quantum entanglement in those systems when multiple degenerate or near-degenerate mechanical oscillators are coupled to a common optical mode. The physical reason behind this obstacle can be explained by the dark-mode effect induced by degenerate mechanical modes coupled to a common cavity mode~\cite{Genes2008,Massel2012,Shkarin2014,Sommer2019,Ockeloen-Korppi2019,Lai2020PRARC}. The dark-mode effect strongly suppresses the generation of entanglement when the system works at a nonzero temperature~\cite{Vitali2007jpa,Genes2008}. Therefore, a natural question is whether the optomechanical entanglement can be generated by breaking the dark-mode effect in the multimode optomechanical systems.

In this paper, we propose the auxiliary-cavity-mode method to generate considerable optomechanical entanglement by breaking the dark-mode effect existed in the two-mechanical-mode optomechanical system. This is realized by introducing an auxiliary cavity mode optomechanically coupled to either one of the two mechanical modes. We find that quantum entanglement between the cavity mode and the two degenerate mechanical modes can be largely enhanced by breaking the dark mode, and the generated optomechanical entanglement is immune to the thermal noise. In particular, the threshold thermal phonon number for preserving the optomechanical entanglement could reach around $400$. In addition, we study quantum entanglement between the two mechanical modes. It is found that the mechanical entanglement is extremely fragile to thermal noise. We also study the universal physical coupling configurations for breaking the dark mode in the four-mode optomechanical network consisting of two cavity modes and two mechanical modes. We find that the creation of optomechanical entanglement depends on the breaking of the dark mode. Our work will provide a different route to the generation of optomechanical entanglement between a cavity mode and two degenerate or near-degenerate mechanical modes.

The rest of this paper is organized as follows. In Sec.~\ref{model}, we introduce the physical model and present the quantum Langevin equations. In Sec.~\ref{entanglement}, we study the enhancement of optomechanical entanglement by breaking the dark mode with the auxiliary-cavity-mode method. In Sec.~\ref{Mechentangle}, we study quantum entanglement between the two mechanical modes. In Sec.~\ref{network}, we study the optomechanical entanglement and the general dark-mode-breaking conditions in a general four-mode optomechanical network in which any two nodes could be coupled with each other. In Sec.~\ref{Discuss}, we present some discussions on the experimental implementation of this scheme, elaborate the mechanism of dark-mode breaking in the absence of the optomechanical-coupling linearization, and establish the dominate function of the auxiliary cavity mode $a_{s}$ in the enhancement of optomechanical entanglement. Finally, we present a brief conclusion in Sec.~\ref{conclu}.

\section{Physical model and equations of motion\label{model}}

\begin{figure}[tbp]
\center
\includegraphics[width=0.47 \textwidth]{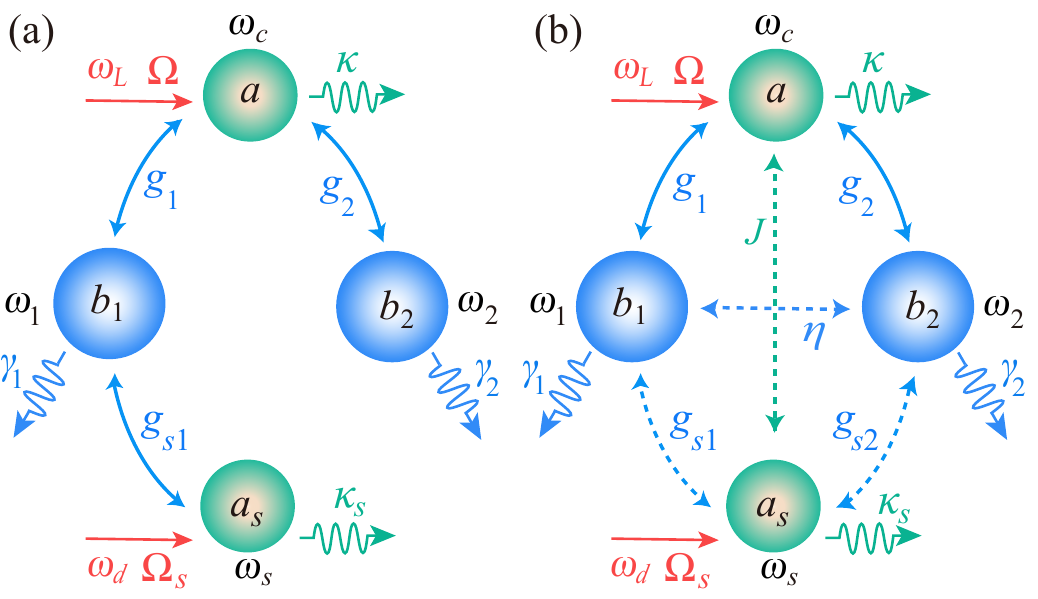}
\caption{(a) Schematic of the four-mode optomechanical system. The intermediate coupling cavity mode $a$ (with resonance frequency $\omega_{c}$, decay rate $\kappa$) is coupled to two mechanical modes $b_{1}$ ($\omega_{1}$, $\gamma_{1}$) and $b_{2}$ ($\omega_{2}$, $\gamma_{2}$) via radiation-pressure interactions (coupling strengths $g_{1}$ and $g_{2}$). The auxiliary cavity mode $a_{s}$ ($\omega_{s}$, $\kappa_{s}$) is optomechanically coupled to the mechanical mode $b_{1}$ (coupling strength $g_{s1}$). Two proper driving fields with driving frequencies $\omega_{L}$ and $\omega_{d}$ and driving amplitudes $\Omega$ and $\Omega_{s}$ are, respectively, introduced to drive the cavity modes $a$ and $a_{s}$. (b) Schematic of a general case of the four-mode optomechanical system, we assume that any two nodes in the system could be coupled with each other. In addition to the notations marked in panel (a), the connections $J$, $\eta$, and $g_{s2}$ stand for the photon hopping, the phonon hopping, and the optomechanical coupling between the auxiliary cavity mode and the mechanical mode $b_{2}$, respectively. Here, for analyzing the dark-mode effect in this optomechanical network, we assume that the two couplings $g_{1}$ and $g_{2}$ (solid lines) always exist, but other four couplings (marked by dashed lines) could be turned on or off on demand.}
\label{Fig1}
\end{figure}

We consider a four-mode optomechanical system consisting of two cavity modes and two mechanical modes, as shown in Fig.~\ref{Fig1}(a). Here, the target system is composed by the intermediate coupling cavity mode $a$ optomechanically coupled to two mechanical modes $b_{1}$ and $b_{2}$ [namely, the modes $a$, $b_{1}$, and $b_{2}$, and the couplings $g_{1}$ and $g_{2}$ in Fig.~\ref{Fig1}(a)]. When the two mechanical modes are degenerate, one of the two mechanical normal modes becomes the dark mode~\cite{Genes2008,Ockeloen-Korppi2019,Lai2020PRARC}. To break the dark mode, we introduce an auxiliary cavity mode $a_{s}$ to couple with the mechanical mode $b_{1}$ via the radiation-pressure interaction. Besides, the driving field with frequency $\omega_{L}$ ($\omega_{d}$) and amplitude $\Omega$ ($\Omega_{s}$) is injected into the cavity mode $a$ ($a_{s}$). In a rotating frame defined by the transformation operator $V(t)=\exp[-i(\omega _{L}a^{\dagger}a+\omega_{d}a_{s}^{\dagger}a_{s})t]$, the Hamiltonian of the system reads
\begin{eqnarray}
H_{I}&=&\Delta_{c}a^{\dagger}a+\Delta_{s}a_{s}^{\dagger}a_{s}+\sum_{l=1,2}[\omega_{l}b_{l}^{\dagger}b_{l}+g_{l}a^{\dagger}a(b_{l}^{\dagger}+b_{l})]\nonumber\\
&&+g_{s1}a_{s}^{\dagger}a_{s}(b_{1}^{\dagger}+b_{1})+(\Omega a^{\dagger}+\Omega _{s}a_{s}^{\dagger }+\mathrm{H.c.}), \label{Hamit1}
\end{eqnarray}
where $a$ $(a^{\dagger})$, $a_{s}$ $(a_{s}^{\dagger})$, and $b_{l=1,2}$ $(b^{\dagger}_{l})$ are the annihilation (creation) operators of the intermediate cavity mode (resonance frequency $\omega_{c}$), the auxiliary cavity mode (resonance frequency $\omega_{s}$), and the $l$th mechanical mode (resonance frequency $\omega_{l}$), respectively. The parameter $\Delta_{c}=\omega_{c}-\omega_{L}$ ($\Delta_{s}=\omega_{s}-\omega_{d}$) is the driving detuning between the cavity mode frequency $\omega_{c}$ ($\omega_{s}$) and its driving field frequency $\omega_{L}$ ($\omega_{d}$). The strength of the single-photon optomechanical coupling  between the cavity mode $a$ ($a_{s}$) and the $l$th mechanical mode $b_{l=1,2}$ ($b_{1}$) is denoted by $g_{l=1,2}$ $(g_{s1})$. Note that the experimental implementation of the physical model with superconducting circuits will be discussed in Sec.~\ref{DiscussA}.

Under the condition of strong driving, one can linearize the dynamics of the system by expanding each operator $o\in\{a, a^{\dagger}, a_{s}, a_{s}^{\dagger}, b_{l}, b^{\dagger}_{l}\}$ as a summation of its steady-state mean value and small fluctuation, i.e., $o=\langle o\rangle_{\text{SS}}+\delta o$. To study the optomechanical entanglement, we introduce the quadrature operators $\delta X_{o}=(\delta o^{\dagger}+\delta o)/\sqrt{2}$ and $\delta Y_{o}=i(\delta o^{\dagger}-\delta o)/\sqrt{2}$ for the mentioned operators: $o=a$,  $a^{\dagger}$, $a_{s}$, $a_{s}^{\dagger}$, $b_{l}$, and $b_{l}^{\dagger}$. In addition, we introduce the corresponding input noise operators $X_{o}^{\text{in}}=(o_{\text{in}}^{\dagger}+o_{\text{in}})/\sqrt{2}$ and $Y_{o}^{\text{in}}=i(o_{\text{in}}^{\dagger}-o_{\text{in}})/\sqrt{2}$ for $o_{\text{in}}=a_{\text{in}}$, $a_{s,\text{in}}$, $b_{1,\text{in}}$, and $b_{2,\text{in}}$. Here, the operators $a_{\textrm{in}}$ $(a^{\dagger}_{\textrm{in}})$, $a_{s,\textrm{in}}$ $(a^{\dagger}_{s,\textrm{in}})$, and $b_{l,\textrm{in}}$ $(b^{\dagger}_{l,\textrm{in}})$ are the noise operators related to the bosonic modes $a$, $a_{s}$, and $b_{l}$, respectively. These noise operators are determined by the nonzero correlation functions~\cite{Gardiner2013}
$\langle a_{\textrm{in}}(t) a_{\textrm{in}}^{\dagger}(t^{\prime})\rangle=\delta(t-t^{\prime})$,
$\langle a_{s,\textrm{in}}(t) a_{s,\textrm{in}}^{\dagger}(t^{\prime})\rangle=\delta(t-t^{\prime})$,
$\langle b_{l,\textrm{in}}(t) b_{l,\textrm{in}}^{\dagger}(t^{\prime})\rangle=(\bar{n}_{l}+1)\delta(t-t^{\prime})$, and
$\langle b_{l,\textrm{in}}^{\dagger}(t) b_{l,\textrm{in}}(t^{\prime})\rangle=\bar{n}_{l}\delta(t-t^{\prime})$
for $l=1,2$, where $\bar{n}_{l=1,2}$ denotes the thermal phonon number associated with the $l$th mechanical mode.

By defining the quadrature fluctuation vector $\mathbf{u}(t)=(\delta{X}_{b_{1}},\delta{Y}_{b_{1}},\delta{X}_{b_{2}},\delta{Y}_{b_{2}},\delta{X}_{a_{s}},\delta{Y}_{a_{s}},\delta{X}_{a},\delta{Y}_{a})^{T}$ and the corresponding input noise operator vector $\mathbf{N}(t)=\sqrt{2}(\sqrt{\gamma_{1}}X_{b_{1}}^{\text{in}},\sqrt{\gamma_{1}}Y_{b_{1}}^{\text{in}},\sqrt{\gamma_{2}}X_{b_{2}}^{\text{in}},\sqrt{\gamma_{2}}Y_{b_{2}}^{\text{in}}
,\sqrt{\kappa_{s}}X_{a_{s}}^{\text{in}},\sqrt{\kappa_{s}}Y_{a_{s}}^{\text{in}},\sqrt{\kappa}X_{a}^{\text{in}},\\\sqrt{\kappa}Y_{a}^{\text{in}})^{T}$, we obtain a compact form of the linearized Langevin equation
\begin{eqnarray}
\mathbf{\dot{u}}(t)=\mathbf{Au}(t)+\mathbf{N}(t),\label{compacteq}
\end{eqnarray}
where the coefficient matrix $\mathbf{A}$ takes the form as
\begin{equation}
\label{coefficient}
\mathbf{A}=\left(
\begin{array}{cccccccc}
-\gamma _{1} & \omega _{1} & 0 & 0 & 0 & 0 & 0 & 0 \\
-\omega _{1} & -\gamma _{1} & 0 & 0 & -2G_{s1} & 0 & -2G_{1} & 0 \\
0 & 0 & -\gamma _{2} & \omega _{2} & 0 & 0 & 0 & 0 \\
0 & 0 & -\omega _{2} & -\gamma _{2} & 0 & 0 & -2G_{2} & 0 \\
0 & 0 & 0 & 0 & -\kappa _{s} & \Delta _{s}^{\prime} & 0 & 0 \\
-2G_{s1} & 0 & 0 & 0 & -\Delta_{s}^{\prime} & -\kappa_{s} & 0 & 0 \\
0 & 0 & 0 & 0 & 0 & 0 & -\kappa  & \Delta _{c}^{\prime} \\
-2G_{1} & 0 & -2G_{2} & 0 & 0 & 0 & -\Delta _{c}^{\prime} & -\kappa
\end{array}
\right).
\end{equation}
Here, $\kappa$, $\kappa_{s}$, and $\gamma_{l=1,2}$ are, respectively, the decay rates of the bosonic modes $a$, $a_{s}$, and $b_{l}$. The parameters $\Delta_{c}^{\prime}=\Delta_{c}+2\sum_{l=1}^{2} g_{l}\text{Re}(\beta_{l})$ and $\Delta_{s}^{\prime}=\Delta_{s}+2g_{s1}\text{Re}(\beta_{1})$ are, respectively, the normalized driving detunings of the cavity modes $a$ and $a_{s}$, with $\text{Re}(\beta_{l=1,2})$ taking the real part of $\beta_{l}$. The parameter $G_{l=1,2}=g_{l}\alpha$ ($G_{s1}=g_{s1}\alpha_{s}$) is the strength of linearized optomechanical coupling between the cavity mode $a$ ($a_{s}$) and the mechanical mode $b_{l=1,2}$ ($b_{1}$). Here, the steady-state mean values $\alpha$, $\alpha_{s}$, and $\beta_{l=1,2}$ are defined by $\alpha=\langle a\rangle_{\text{SS}}$, $\alpha_{s}=\langle a_{s}\rangle_{\text{SS}}$, and $\beta_{l=1,2}=\langle b_{l}\rangle_{\text{SS}}$. The value of these variables can be  obtained by calculating the steady-state mean-value equations. Note that the steady-state values of $\alpha$ and $\alpha_{s}$ could be real by choosing proper phases of the driving amplitudes $\Omega$ and $\Omega_{s}$, then the linearized optomechanical-coupling strengths $G_{l=1,2}$ and $G_{s1}$ are real accordingly. The stability of this linearized optomechanical system can be analyzed with the Routh-Hurwitz criterion~\cite{Gradstein2014}. The parameters used in the following calculations satisfy the stability conditions, i.e., confirming that the real parts of all the eigenvalues of the coefficient matrix $\mathbf{A}$ are negative.

For studying the steady-state optomechanical entanglement between the cavity mode $a$ and the mechanical modes $b_{l=1,2}$, we introduce the covariance matrix $\mathbf{V}$ of this system, defined by the matrix elements
\begin{equation}
\mathbf{V}_{ij}=\frac{1}{2}[\langle \mathbf{u}_{i}(\infty) \mathbf{u}_{j}(\infty) \rangle +\langle \mathbf{u}_{j}( \infty) \mathbf{u}_{i}(\infty )\rangle], \hspace{0.5 cm}i,j=1\text{-}8. \label{covariance}
\end{equation}
This covariance matrix $\mathbf{V}$ satisfies the Lyapunov equation~\cite{Vitali2007prl}
\begin{equation}
\mathbf{A}\mathbf{V}+\mathbf{V}\mathbf{A}^{T}=-\mathbf{Q}, \label{Lyapunov}
\end{equation}
where $\mathbf{Q}=\mathrm{diag} \{(2\bar{n}_{1}+1)\gamma_{1},(2\bar{n}_{1}+1)\gamma_{1},(2\bar{n}_{2}+1)\gamma_{2},(2\bar{n}_{2}+1)\gamma_{2},\kappa_{s},\kappa_{s},\kappa,\kappa\}$ is the diffusion matrix.

\begin{figure}[tbp]
\centering
\includegraphics[width=0.48 \textwidth]{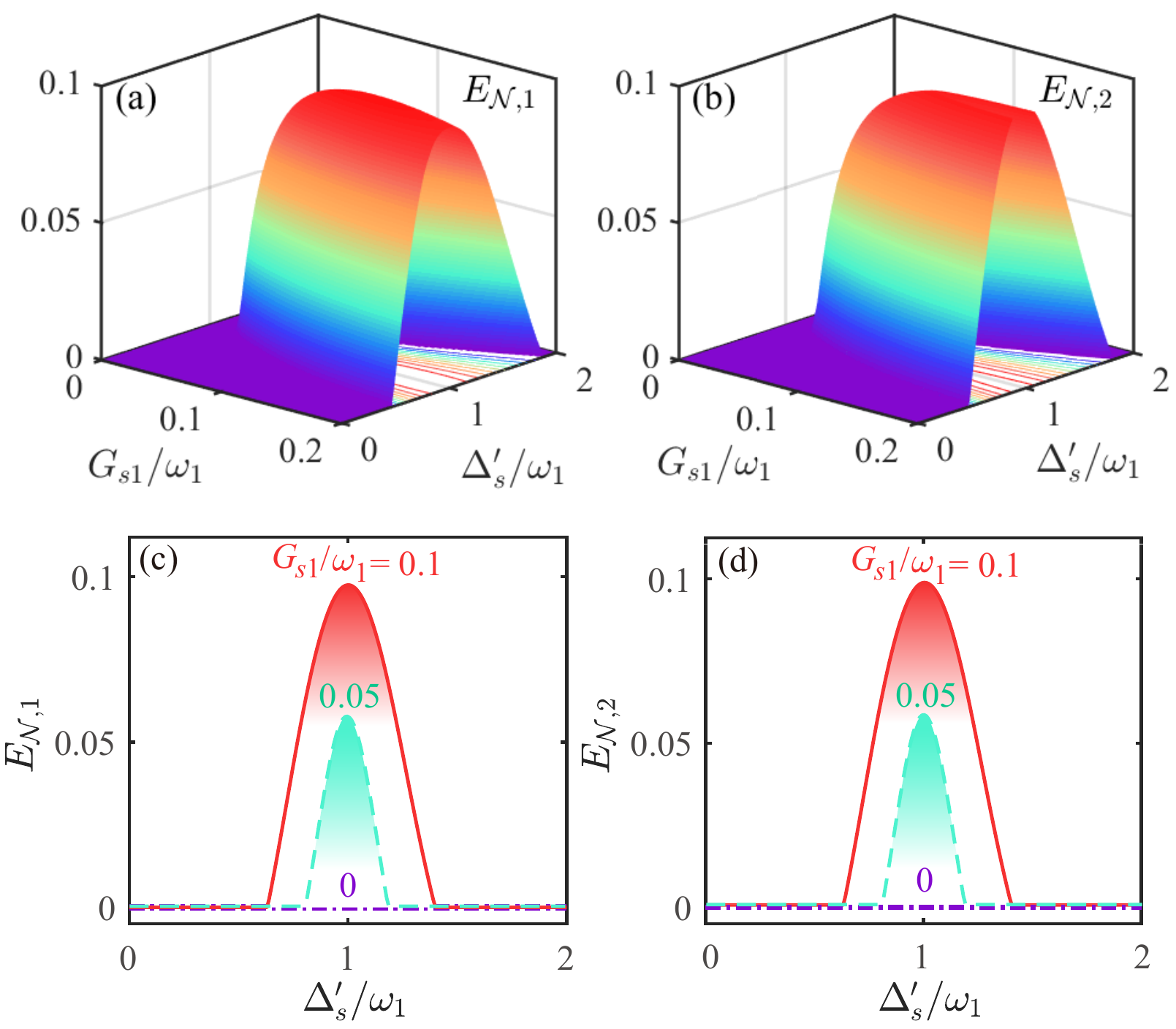}
\caption{(Color online) Logarithmic negativities (a) $E_{\mathcal{N},1}$ and (b) $E_{\mathcal{N},2}$ versus the scaled driving detuning $\Delta_{s}^{\prime}/\omega_{1}$ and the scaled optomechanical coupling strength $G_{s1}/\omega_{1}$. (c) $E_{\mathcal{N},1}$ and (d) $E_{\mathcal{N},2}$ versus $\Delta_{s}^{\prime}/\omega_{1}$ in both the dark-mode-unbreaking ($G_{s1}/\omega_{1}=0$, dash-dot blue curves) and -breaking ($G_{s1}/\omega_{1}=0.05$, dashed cyan curves, and $G_{s1}/\omega_{1}=0.1$, solid red curves) cases. Other parameters used are $\omega_{2}=\omega_{1}$, $\gamma_{1}/\omega_{1}=\gamma_{2}/\omega_{1}=10^{-5}$, $G_{1}/\omega_{1}=G_{2}/\omega_{1}=0.15$, $\Delta_{c}^{\prime}/\omega_{1}=1$, $\kappa/\omega_{1}=\kappa_{s}/\omega_{1}=0.1$, and $\bar{n}_{1}=\bar{n}_{2}=100$.}
\label{Fig2}
\end{figure}

\section{Optomechanical entanglement \label{entanglement}}

In this section, we study the enhancement of optomechanical entanglement by breaking the mechanical dark mode with the auxiliary-cavity-mode method. Concretely, we calculate the logarithmic negativity between the cavity mode $a$ and the mechanical mode $b_{l=1,2}$ in both the presence and absence of the auxiliary cavity mode.

\subsection{Generation of optomechanical entanglement}

To describe quantum entanglement between the cavity field and each mechanical mode, we first introduce the definition of the logarithmic negativity for a system consisting of two bosonic modes $A$ and $B$ in a Gaussian state. We denote the covariance matrix of the two-mode system as $\mathcal{V}$, which can be expressed as
\begin{equation}
\mathcal{V}=\left(
\begin{array}{cc}
\mathcal{V}_{A} & \mathcal{V}_{AB} \\
\mathcal{V}_{AB}^{T} & \mathcal{V}_{B}
\end{array}
\right).\label{reduce}
\end{equation}
Here $\mathcal{V}_{A}$, $\mathcal{V}_{B}$, and $\mathcal{V}_{AB}$ are $2\times2$ block matrices, which are related to the mode $A$, the mode $B$, and the two-mode correlation, respectively.  The logarithmic negativity~\cite{Vidal2002,Plenio2005} is defined by
\begin{equation}
E_{\mathcal{N}}=\max [0,-\ln(2\varsigma^{-})],\label{negativity}
\end{equation}
where
\begin{equation}
\varsigma^{-}=\frac{1}{\sqrt{2}}\left[\Sigma\left(\mathcal{V}\right)-\sqrt{\Sigma\left(\mathcal{V}\right)^{2}-4\det\mathcal{V}}\right]^{1/2}
\end{equation}
is the smallest eigenvalue of the covariance matrix $\mathcal{V}$, with $\Sigma\left(\mathcal{V}\right)=\det \mathcal{V}_{A}+\det \mathcal{V}_{B}-2\det \mathcal{V}_{AB}$. From Eq.~(\ref{negativity}) we can find that entanglement exists in this system only when $\varsigma^{-}<1/2$, which is equivalent to Simon's entanglement criterion~\cite{Simon2000}.

In the following, we introduce the logarithmic negativity $E_{\mathcal{N},l=1,2}$ to measure optomechanical entanglement between the cavity mode $a$ and the $l$th mechanical mode $b_{l=1,2}$.  Accordingly, $\mathcal{V}_{B}$, $\mathcal{V}_{A}$, and $\mathcal{V}_{AB}$ should be replaced by the block matrices of the cavity mode $a$, the $l$th mechanical mode $b_{l=1,2}$, and the optomechanical correlations, respectively. Concretely, when one calculates the entanglement between the cavity mode $a$ and the mechanical mode $b_{1}$ ($b_{2}$),  the reduced matrix $\mathcal{V}$ of modes $a$ and $b_{1}$ ($b_{2}$) can be obtained by tracing out the rows and columns of modes $a_{s}$ and $b_{2}$ ($b_{1}$) in the covariance matrix $\mathbf{V}$ given in Eq.~(\ref{covariance}).

\begin{figure}[tbp]
\centering
\includegraphics[width=0.46 \textwidth]{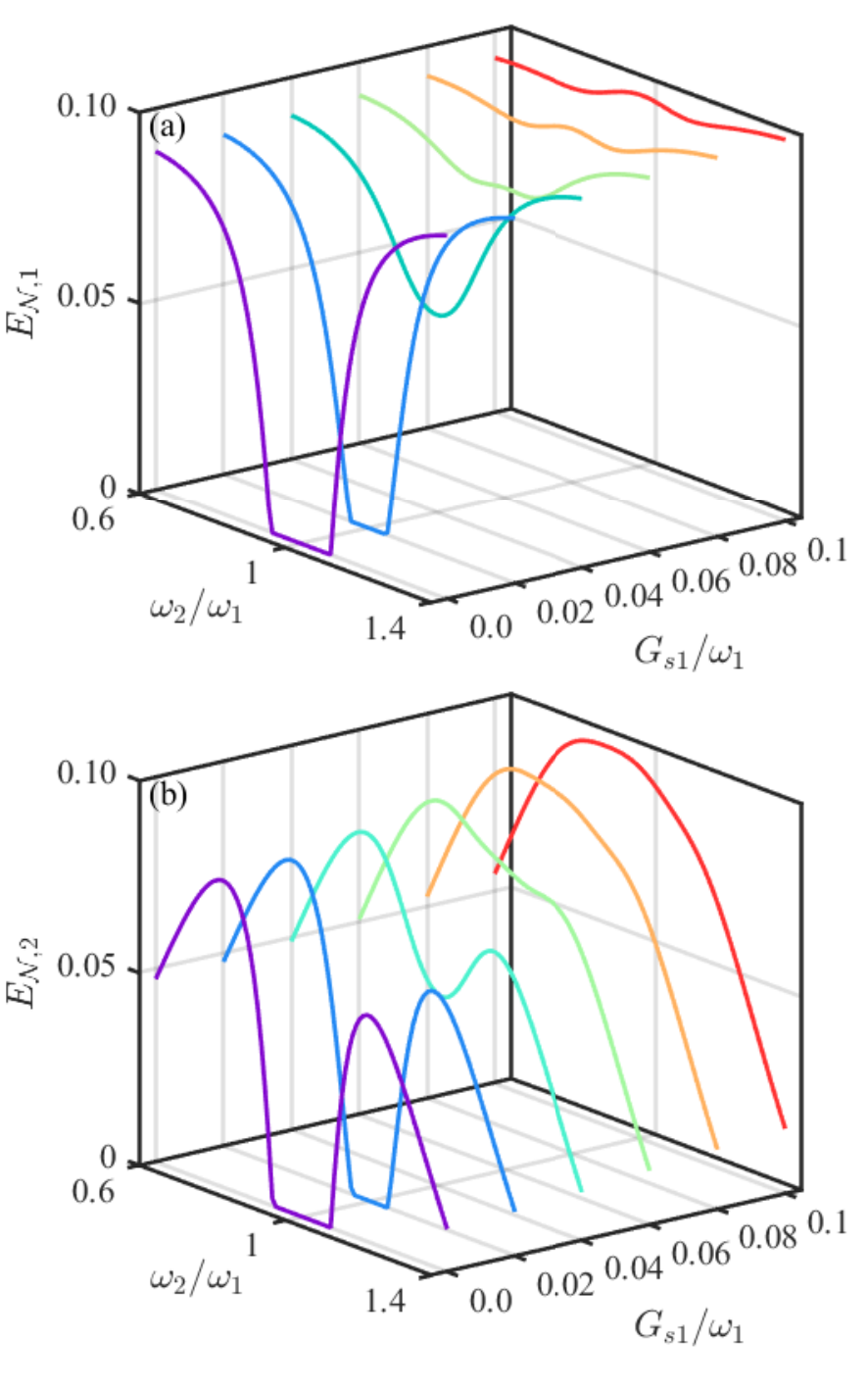}
\caption{(Color online) Logarithmic negativities (a) $E_{\mathcal{N},1}$ and (b) $E_{\mathcal{N},2}$ versus the the frequency ratio $\omega_{2}/\omega_{1}$ when the optimal driving $\Delta_{c}^{\prime}=\Delta_{s}^{\prime}=\omega_{1}$ and the scaled optomechanical coupling strength takes various values $G_{s1}/\omega_{1}=0$, $0.02$, $0.04$, $0.06$, $0.08$, and $0.1$. Other parameters are $\gamma_{1}/\omega_{1}=\gamma_{2}/\omega_{1}=10^{-5}$, $G_{1}/\omega_{1}=G_{2}/\omega_{1}=0.15$, $\kappa/\omega_{1}=\kappa_{s}/\omega_{1}=0.1$, and $\bar{n}_{1}=\bar{n}_{2}=100$.}
\label{Fig3}
\end{figure}

The dark mode formed by the two degenerate mechanical modes coupled to a common cavity mode is an obstacle against the generation of optomechanical entanglement when the system works at a nonzero temperature~\cite{Genes2008,Lai2022}. A natural question is whether we can break the dark-mode effect to enhance the optomechanical entanglement in this system. To address this concern, we investigate the entanglement measure when the auxiliary cavity mode exists or does not exist. In Figs.~\ref{Fig2}(a) and~\ref{Fig2}(b), the logarithmic negativities $E_{\mathcal{N},1}$ and $E_{\mathcal{N},2}$ are plotted as functions of the scaled driving detuning $\Delta_{s}^{\prime}/\omega_{1}$ and the scaled optomechanical coupling strength $G_{s1}/\omega_{1}$ associated with the auxiliary cavity mode. Here, $E_{\mathcal{N},1}$ ($E_{\mathcal{N},2}$) is used to characterize the optomechanical entanglement between the cavity mode $a$ and the mechanical mode $b_{1}$ ($b_{2}$). We can see that the logarithmic negativity reaches the maximum value around the optimal driving detuning $\Delta_{s}^{\prime}=\omega_{1}$. When the coupling strength $G_{s1}/\omega_{1}$ is close to $0$, the logarithmic negativity becomes zero. This feature can be seen more clearly in Figs.~\ref{Fig2}(c) and~\ref{Fig2}(d). Here, we see that the cavity mode $a$ and the mechanical modes are uncorrelated in the absence of the auxiliary cavity mode ($E_{\mathcal{N},1}=E_{\mathcal{N},2}=0$, the lower horizontal dash-dot purple curves), but they are entangled in the presence of the auxiliary cavity mode ($E_{\mathcal{N},1}\approx E_{\mathcal{N},2}\approx0.1$, the upper solid red curves). The results indicate that the auxiliary cavity mode $a_{s}$ plays a very important role in the enhancement of the optomechanical entanglement.

The underlying physics of this entanglement enhancement can be explained using the dark-mode physical mechanism~\cite{Genes2008,Lai2022}. When the auxiliary cavity mode is absent, the system is reduced to two degenerate mechanical modes $b_{1}$ and $b_{2}$ coupled to a common cavity mode $a$, then there exists a dark mode~\cite{Genes2008,Lai2020PRARC,Liu2022,Huang2021}. This dark mode is decoupled from the cavity mode and the mechanical bright mode, and hence the dark mode cannot be cooled. Further, the thermal excitations stored in the dark mode will destroy all the quantum effects, including optomechanical entanglement~\cite{Lai2022}. However, in the presence of the auxiliary cavity mode, the dark-mode effect is broken, then both the mechanical modes can be cooled to their ground states. In this case, quantum entanglement can be created by the optomechanical interaction. Thus, the auxiliary-cavity-mode method not only provides the physical origin for breaking the dark mode, but also shows a clear perspective for creating dark-mode-immune quantum resources in coupled multiple-mechanical-resonator systems. The phenomenon of this entanglement enhancement assisted by the ground-state cooling can be confirmed based on the optical detuning effect. As shown in Figs.~\ref{Fig2}(a) and~\ref{Fig2}(b), the maximal entanglement appears around $\Delta_{s}^{\prime}=\omega_{1}$. This is because the photon-phonon exchanging becomes the most efficient at the red-sideband resonance, and then the optimal cooling occurs in this case. In addition, we see from Figs.~\ref{Fig2}(a) and~\ref{Fig2}(b) that the logarithmic negativities $E_{\mathcal{N},l=1,2}$ increase with the increase of the scaled optomechanical coupling strength $G_{s1}/\omega_{1}$. This phenomenon can be understood by analyzing the role of the dark-mode effect. Although the increase of the coupling strength between the auxiliary cavity mode and the mechanical mode will increase the overall mixing degree of the original optomechanical system, the dark-mode effect is the dominate factor affecting the generation of the optomechanical entanglement, thus the optomechanical entanglement increases with the increase of the coupling strength $G_{s1}/\omega_{1}$, which is used to break the dark mode.

Since the dark mode appears only theoretically in the two-degenerate-mechanical-mode case, it becomes an interesting topic to analyze the influence of the frequency mismatch between the two mechanical modes on the generation of optomechanical entanglement. In Figs.~\ref{Fig3}(a) and~\ref{Fig3}(b) we plot the logarithmic negativities $E_{\mathcal{N},1}$ and $E_{\mathcal{N},2}$ as functions of the frequency ratio $\omega_{2}/\omega_{1}$ in both the dark-mode-unbreaking ($G_{s1}/\omega_{1}=0$) and -breaking ($G_{s1}/\omega_{1}=0.02$, $0.04$, $0.06$, $0.08$, and $0.1$) cases. Here, we can see that, although the dark mode exists theoretically only in the degenerate-mechanical-mode case (i.e., $\omega_{1}$=$\omega_{2}$), the dark-mode effect actually works for a wider detuning range in the near-degenerate-mechanical-mode case (see the valley area of blue curves in Fig.~\ref{Fig3}). Due to the influence of this dark-mode effect, the logarithmic negativities $E_{\mathcal{N},1}$ and $E_{\mathcal{N},2}$ are greatly suppressed in this valley area.  When the auxiliary cavity mode is introduced into this system ($G_{s1}/\omega_{1}\neq0$), the optomechanical entanglement for the degenerate and near-degenerate mechanical modes becomes feasible, because the dark mode is broken (see the other curves in Fig.~\ref{Fig3}). In addition, we \textcolor{blue}{find} that the optomechanical entanglement $E_{\mathcal{N},1}$ and $E_{\mathcal{N},2}$ have different dependence features with the frequency ratio $\omega_{2}/\omega_{1}$. In Fig.~\ref{Fig3} we chose the mechanical frequency $\omega_{1}$ as the scale unit and consider the resonance case $\Delta_{c}^{\prime}=\Delta_{s}^{\prime}=\omega_{1}$. Therefore, the coupling between the cavity mode $a$ and the mechanical mode $b_{1}$ is resonant. Different from the mechanical mode $b_{1}$, the mode $b_{2}$ will be coupled unresonantly with the mode $a$ when $\omega_{2}/\omega_{1}\neq1$. When the modes $a$ and $b_{2}$ are of far-off-resonance, the optomechanical entanglement $E_{\mathcal{N},2}$ between the cavity mode $a$ and  the mechanical mode $b_{2}$ is not as good as that of the  optomechanical entanglement $E_{\mathcal{N},1}$. This result is consistent with the analysis obtained in Fig.~\ref{Fig2} (entanglement enhancement assisted by the ground-state cooling at the optimal drive detuning).

\subsection{Thermal-noise-resistant optomechanical entanglement}

\begin{figure}[tbp]
\centering
\includegraphics[width=0.46 \textwidth]{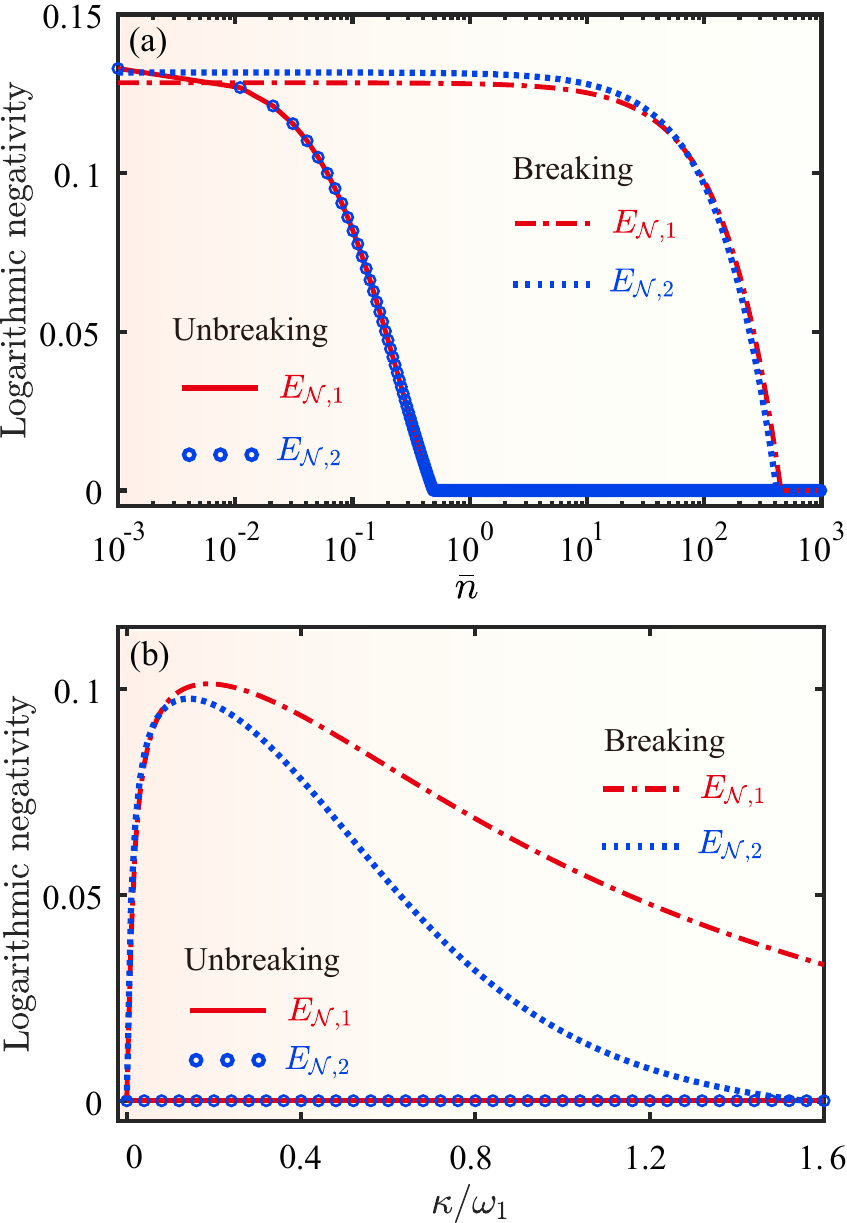}
\caption{(Color online) Logarithmic negativities $E_{\mathcal{N},1}$ and $E_{\mathcal{N},2}$ versus (a) the thermal phonon numbers $\bar{n}$ or (b) the scaled decay rate $\kappa/\omega_{1}$ in both the dark-mode-unbreaking (solid curves and circles) and -breaking (dash-dotted curves and dotted curves) cases. Note that the scaled decay rate $\kappa/\omega_{1}=0.1$ is used in panel (a), and the thermal phonon numbers $\bar{n}=100$ is used in panel (b). Other parameters used are $\omega_{2}=\omega_{1}$, $\bar{n}_{1}=\bar{n}_{2}=\bar{n}$, $\gamma_{1}/\omega_{1}=\gamma_{2}/\omega_{1}=10^{-5}$, $G_{1}/\omega_{1}=G_{2}/\omega_{1}=0.15$, $G_{s1}/\omega_{1}=0.1$, $\Delta_{c}^{\prime}/\omega_{1}=\Delta_{s}^{\prime}/\omega_{1}=1$, and $\kappa_{s}/\omega_{1}=0.1$.}
\label{Fig4}
\end{figure}

When the dark-mode effect appears ($G_{s1}=0$), the optomecanical entanglement between the cavity mode and the mechanical modes is extremely fragile to the thermal noise. To study whether the auxiliary-cavity-mode method can generate robust quantum entanglement against the thermal noise, we plot the dependence of the quantum entanglement on the thermal mechanical excitation number $\bar{n}_{1}=\bar{n}_{2}=\bar{n}$ in both the dark-mode-unbreaking and -breaking cases. As shown in Fig.~\ref{Fig4}(a), the logarithmic negativities $E_{\mathcal{N},1}$ and $E_{\mathcal{N},2}$ emerge only when the system is approximately in the quantum ground state ($\bar{n}_{l=1,2}\ll1$) (see the horizontal axis ranges from $10^{-3}$ to $1$) in the dark-mode-unbreaking case. This means that there is no optomechanical entanglement between the cavity mode and the mechanical modes when the mechanical modes are not precooled to their quantum ground states. In the dark-mode-breaking case, differently, the logarithmic negativities $E_{\mathcal{N},1}$ and $E_{\mathcal{N},2}$ can still exist even when the thermal phonons in the two mechanical modes are about $\bar{n}_{l=1,2}=450$ (see the horizontal axis ranges from $10^{2}$ to $10^{3}$), which is about three orders of magnitude larger than that in the dark-mode-unbreaking case.

The physical reason for this phenomenon can be understood as follows. When this system works at a nonzero temperature, the thermal excitation energy stored in the dark mode cannot be extracted through the cavity mode, and hence the generation of optomechanical entanglement is largely suppressed by the thermal noise. However, when the auxiliary cavity mode is introduced, the dark-mode effect is broken. As a result, the mechanical modes can be cooled to approach their ground states, and quantum entanglement can be generated by the optomechanical interaction. This dark-mode-breaking mechanism makes the optomechanical entanglement switchable from extremely fragile to quite robust against the thermal noise.

To investigate the influence of the sideband-resolution condition on the entanglement generation, we plot the logarithmic negativities $E_{\mathcal{N},1}$ and $E_{\mathcal{N},2}$ as functions of the scaled decay rate $\kappa/\omega_{1}$ in both the dark-mode-unbreaking (solid curves) and -breaking (dashed curves) cases. Figure~\ref{Fig4}(b) shows that the logarithmic negativities $E_{\mathcal{N},1}$ and $E_{\mathcal{N},2}$ are destroyed by the thermal noise when the dark mode exists (see the lower solid curves and circles). However, the optomechanical entanglement between the cavity mode and two mechanical modes can be generated when the dark mode is broken (see the upper dash-dotted curves and dotted curves). Moreover, the maximal optomechanical entanglement is located around $\kappa/\omega_{1}=0.2$, which is consistent with the optimal condition for resolved-sideband cooling~\cite{Wilson-Rae2007,Marquardt2007}.

\section{Quantum entanglement between the two mechanical modes\label{Mechentangle}}

In Sec.~\ref{entanglement}, we have investigated how to break the dark-mode effect and generate optomechanical entanglement between the cavity mode and two mechanical modes by using the auxiliary-cavity-mode method. Since the investigated system consists of two mechanical modes optomechanically coupled to a common cavity mode, a natural question arising is how does the dark-mode-breaking effect affect quantum entanglement between the two mechanical modes. To answer this question, we next study quantum entanglement between the two mechanical modes.

\begin{figure}[tbp]
\centering
\includegraphics[width=0.46 \textwidth]{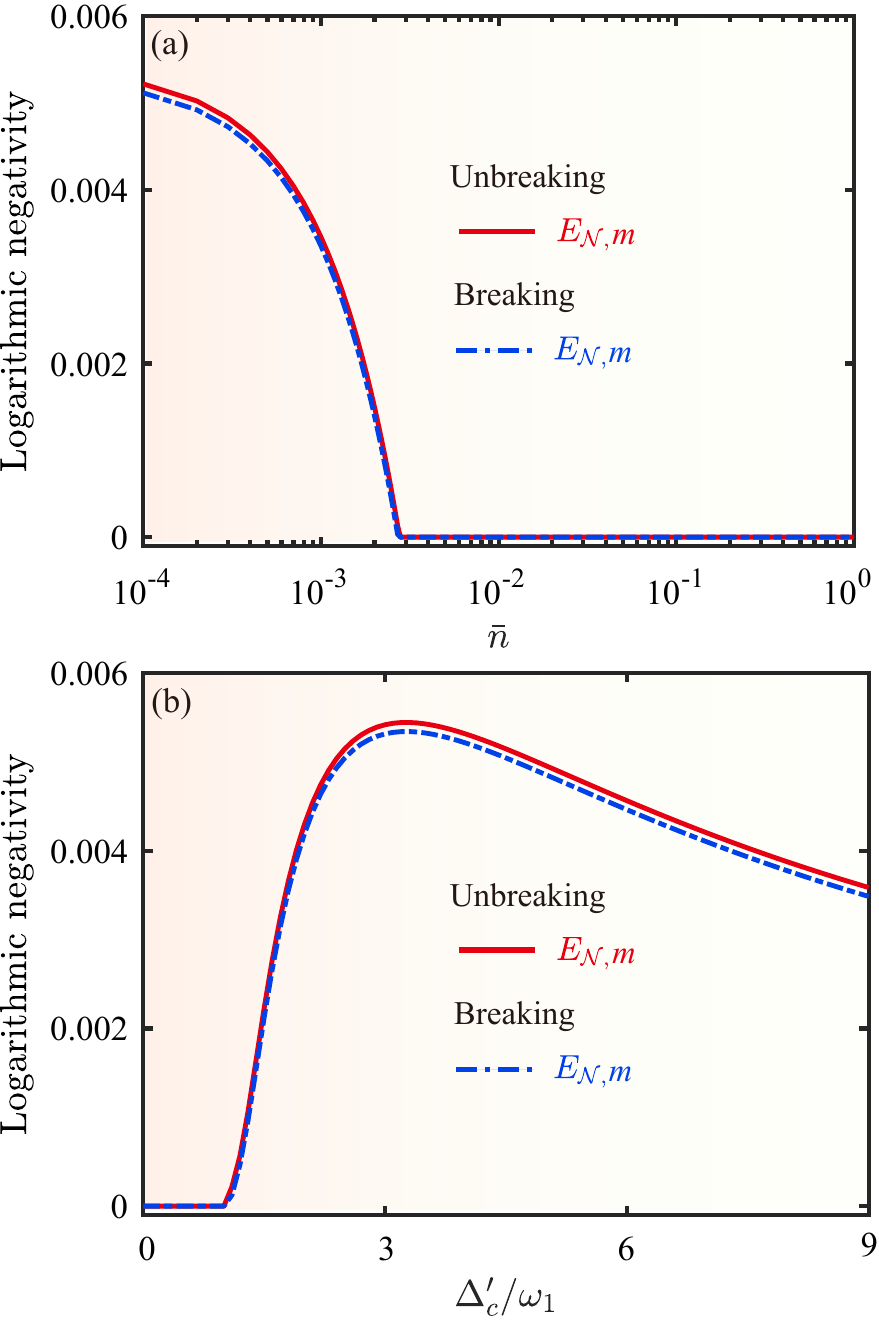}
\caption{Logarithmic negativity $E_{\mathcal{N},m}$ versus (a) the thermal phonon numbers $\bar{n}$ or (b) the scaled driving detuning $\Delta_{c}^{\prime}$ in both the dark-mode-unbreaking (red solid curves) and -breaking (blue dash-dotted curves) cases. Note that the scaled decay rate $\Delta_{c}^{\prime}/\omega_{1}=3$ is used in panel (a), and the thermal phonon numbers $\bar{n}=0$ is used in panel (b). Other parameters used are $\omega_{2}=\omega_{1}$, $\bar{n}_{1}=\bar{n}_{2}=\bar{n}$, $\gamma_{1}/\omega_{1}=\gamma_{2}/\omega_{1}=0.1$, $G_{1}/\omega_{1}=G_{2}/\omega_{1}=0.15$, $G_{s1}/\omega_{1}=0.05$, $\Delta_{s}^{\prime}/\omega_{1}=5$, and $\kappa/\omega_{1}=\kappa_{s}/\omega_{1}=0.1$.}
\label{Fig5}
\end{figure}

In Fig.~\ref{Fig5}(a), we study the resistance of the mechanical entanglement to the thermal effect by plotting the logarithmic negativities $E_{\mathcal{N},m}$ as functions of the thermal excitation number $\bar{n}$ in both the dark-mode-unbreaking (see the red solid curve) and -breaking (see the blue dash-dotted curve) regimes. We find that the logarithmic negativities $E_{\mathcal{N},m}$ are zero in both the dark-mode-unbreaking and -breaking regimes when the thermal excitation number $\bar{n}\geq0.002$, which means that the mechanical entanglement is extremely fragile with respect to thermal noise. These results are fullly consistent with the stationary entanglement between two movable mirrors in a Fabry-Perot cavity~\cite{Vitali2007jpa}. In the low-temperature limit, $\bar{n}\ll1$, the logarithmic negativity has a very small value, and it rapidly approaches zero with the increase of environmental temperature. In addition, even in the almost vacuum-bath case, the mechanical entanglement is also negligible small. This is because there is no direct interaction between the two mechanical modes, and then it is very hard to create the mixed state entanglement between the two mechanical modes. This indicates that the mechanical entanglement should be much smaller than the optomechanical entanglement between the optical mode and the mechanical mode.

In Fig.~\ref{Fig5}(b), we plot the logarithmic negativities $E_{\mathcal{N},m}$ as functions of the scaled driving detuning $\Delta_{c}^{\prime}/\omega_{1}$ in both the dark-mode-unbreaking (the red solid curve) and -breaking (the blue dash-dotted curve) cases.
Here we see that the logarithmic negativities $E_{\mathcal{N},m}$ increase with the increase of driving detuning $\Delta_{c}^{\prime}/\omega_{1}$ in the  initial detuning region $\Delta_{c}^{\prime}/\omega_{1}\approx1\text{-}3$, and reach the maximum at an intermediate detuning region $\Delta_{c}^{\prime}/\omega_{1}\approx3\text{-}3.5$, then the logarithmic negativities decrease in the detuning region $\Delta_{c}^{\prime}/\omega_{1}>3.5$. In Sec.~\ref{entanglement}, we find that optomechanical entanglement between the cavity mode and the mechanical mode can be greatly enhanced when the dark mode is broken, but in Fig.~\ref{Fig5}(b) we find that the mechanical entanglement does not increase when the dark mode is broken. This phenomenon can be understood from the entanglement in the parameters $\bar{n}_{1}=\bar{n}_{2}=0$. In this case, there is no thermal-excitation energy in the dark mode formed by these mechanical modes. Therefore, the auxiliary cavity mode cannot play an important role in the extraction of the thermal-excitation energy from the dark mode. Namely, the mechanical entanglement does not increase even if the dark mode is broken. In addition, the optimal driving detuning for the peak mechanical entanglement appears around $\Delta_{c}^{\prime}/\omega_{1}\approx3$. This is because in the vacuum-bath case, the main reason for the entanglement generation should be the counter-rotating coupling term, which becomes important around the optical driving detuning in this case.

\section{Optomechanical entanglement in the network-coupled four-mode optomechanical system \label{network}}

In the above sections, we showed the generation of optomechanical entanglement by introducing an auxiliary cavity mode $a_{s}$ optomechanically coupled to the mechanical mode $b_{1}$. However, in practical optomechanical networks, the interactions among these bosonic modes are more complicated~\cite{Heinrich2011}. So it is necessary to study the optomechanical entanglement in a general four-mode optomechanical system. Here, we consider a network-coupled four-mode optomechanical system in which any two nodes could be coupled to each other [see Fig.~\ref{Fig1}(b)]. In a rotating frame defined by the operator $V(t)=\exp[-i(\omega_{L}a^{\dagger}a+\omega_{d}a_{s}^{\dagger}a_{s})t]$ with $\omega_{L}=\omega_{d}$, the Hamiltonian of this four-mode optomechanical network is given by
\begin{eqnarray}
H_{I}&=&\Delta_{c}a^{\dagger}a+\Delta_{s}a_{s}^{\dagger}a_{s}+J(a^{\dagger}a_{s}+a_{s}^{\dagger}a)+\eta(b_{1}^{\dagger}b_{2}+b_{2}^{\dagger}b_{1})\nonumber\\
&&+\sum_{l=1,2}[\omega_{l}b_{l}^{\dagger}b_{l}+g_{l}a^{\dagger}a(b_{l}^{\dagger}+b_{l})+g_{sl}a_{s}^{\dagger}a_{s}(b_{l}^{\dagger}+b_{l})]\nonumber\\
&&+(\Omega a^{\dagger}+\Omega_{s}a_{s}^{\dagger }+\mathrm{H.c.}), \label{Hamitfull}
\end{eqnarray}
where the parameters $g_{s2}$, $J$, and $\eta$ are the optomechanical coupling strength, the photon-hopping coupling strength, and the phonon-hopping coupling strength, respectively. Other parameters in Eq.~(\ref{Hamitfull}) have been defined in Eq.~(\ref{Hamit1}).

Based on the Hamiltonian~(\ref{Hamitfull}), we obtain the linearized Langevin equations. According to the similar procedure performed in Sec.~\ref{model},
we can obtain a compact form of the linearized Langevin equations
\begin{eqnarray}
\mathbf{\dot{u}}(t)=\mathbf{\tilde{A}u}(t)+\mathbf{N}(t),\label{MatrixLeq}
\end{eqnarray}
where $\mathbf{u}(t)$ and $\mathbf{N}(t)$ have been defined in Sec.~\ref{model}, and the coefficient matrix $\mathbf{\tilde{A}}$ takes the form as
\begin{equation}
\label{coefficient}
\mathbf{\tilde{A}}=\left(
\begin{array}{cccccccc}
-\gamma _{1} & \omega _{1} & 0 & \eta & 0 & 0 & 0 & 0 \\
-\omega _{1} & -\gamma _{1} & -\eta & 0 & -2G_{s1} & 0 & -2G_{1} & 0 \\
0 & \eta & -\gamma _{2} & \omega _{2} & 0 & 0 & 0 & 0 \\
-\eta & 0 & -\omega _{2} & -\gamma _{2} & -2G_{s2} & 0 & -2G_{2} & 0 \\
0 & 0 & 0 & 0 & -\kappa _{s} & \Delta _{s}^{\prime\prime} & 0 & J \\
-2G_{s1} & 0 & -2G_{s2} & 0 & -\Delta _{s}^{\prime\prime} & -\kappa_{s} & -J & 0 \\
0 & 0 & 0 & 0 & 0 & J & -\kappa  & \Delta _{c}^{\prime} \\
-2G_{1} & 0 & -2G_{2} & 0 & -J & 0 & -\Delta_{c}^{\prime} & -\kappa
\end{array}
\right).
\end{equation}
Here $\Delta_{s}^{\prime\prime}=\Delta_{s}+2g_{s1}\text{Re}(\beta_{1}) +2g_{s2}\text{Re}(\beta_{2})$ is the driving detuning of cavity mode $a_{s}$. Note that the parameters $\Delta_{c}^{\prime}$, $G_{l}$, and $G_{sl(l=1,2)}$ have the same form as the parameters defined in Sec.~\ref{model}, and the steady-state mean values $\alpha$, $\alpha_{s}$, $\beta_{1}$, and $\beta_{2}$ are obtained by calculating the steady-state mean-value equations.

\begin{figure}[tbp]
\centering
\includegraphics[width=0.48 \textwidth]{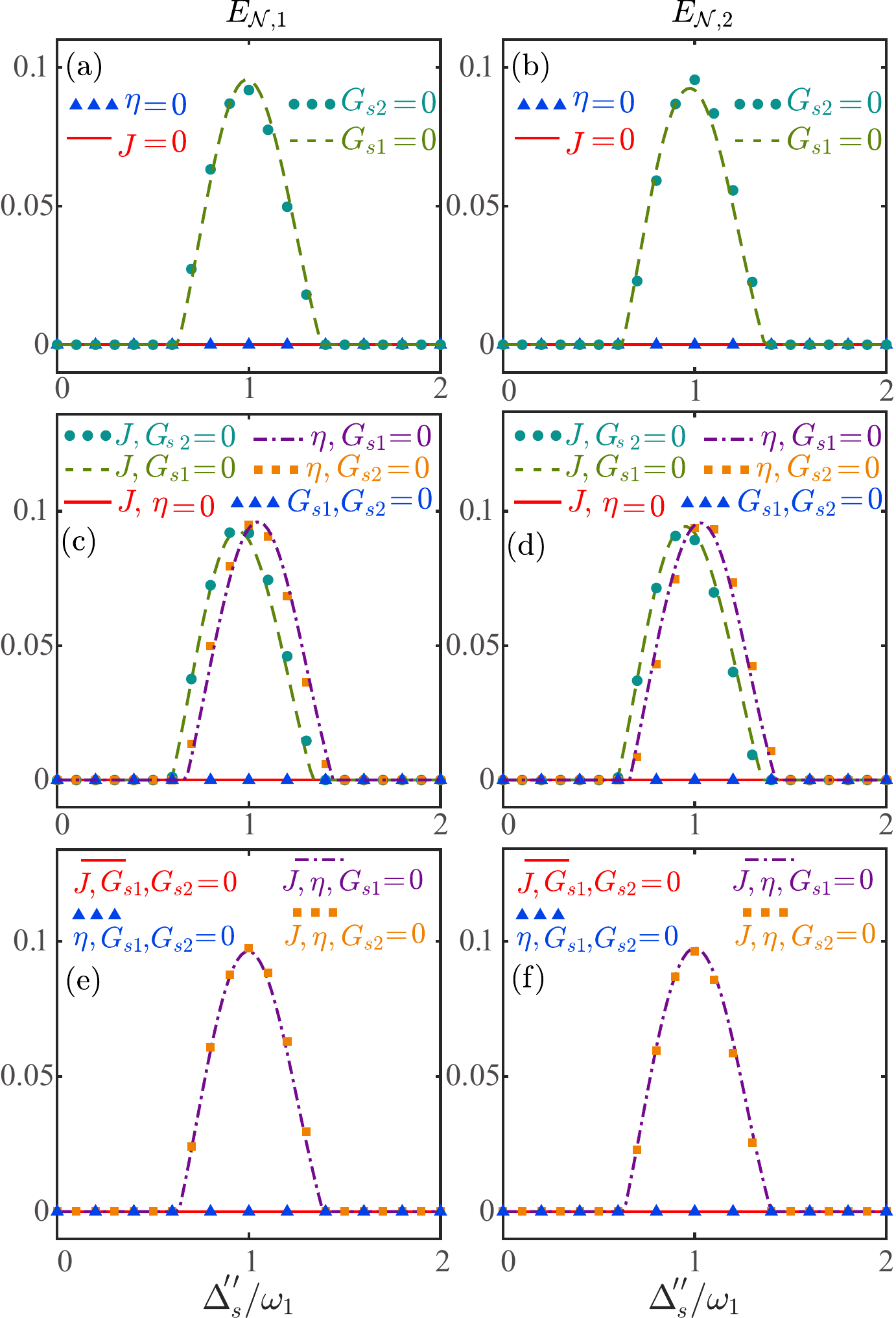}
\caption{(Color online) Logarithmic negativities $E_{\mathcal{N},1}$ and $E_{\mathcal{N},2}$ versus the scaled driving detuning $\Delta_{s}^{\prime\prime}/\omega_{1}$ in various coupling configurations. Panels (a) and (b) indicate the cases where one coupling channel is switched off, i.e., $J=0$ (red solid curves), $\eta=0$ (blue triangles), $G_{s1}=0$ (green dashed curves), and $G_{s2}=0$ (cyan circles). Panels (c) and (d) correspond to the cases where two coupling channels are switched off, i.e.,  $J=\eta=0$ (red solid curves), $G_{s1}=G_{s2}=0$ (blue triangles), $J=G_{s1}=0$ (green dashed curves), $J=G_{s2}=0$ (cyan circles), $\eta=G_{s1}=0$ (purple dash-dotted curves), and $\eta=G_{s2}=0$ (brown squares). Panels (e) and (f) are related to the three-coupling-channel-closed cases: $J=G_{s1}=G_{s2}=0$ (red solid curves), $\eta=G_{s1}=G_{s2}=0$ (blue triangles), $J=\eta=G_{s1}=0$ (purple dash-dotted curves), and $J=\eta=G_{s2}=0$ (brown squares). Other parameters used are taken as $\omega_{2}/\omega_{1}=1$, $\gamma_{1}/\omega_{1}=\gamma_{2}/\omega_{1}=10^{-5}$, $\kappa/\omega_{1}=\kappa_{s}/\omega_{1}=0.1$, $J/\omega_{1}=\eta/\omega_{1}=0.05$, $\Delta_{c}^{\prime}/\omega_{1}=\Delta_{s}^{\prime\prime}/\omega_{1}=1$, $G_{1}/\omega_{1}=G_{2}/\omega_{1}=0.15$, $G_{s1}/\omega_{1}=G_{s2}/\omega_{1}=0.1$, and $\bar{n}_{1}=\bar{n}_{2}=100$. Note that when the values of $J$, $\eta$, $G_{s1}$, and $G_{s2}$ are equal to $0$, the corresponding coupling channels are switched off.}
\label{Fig6}
\end{figure}

The dark-mode effect in this network-coupled four-mode optomechanical system can be clarified by deriving the parameter conditions for the existence of the dark mode. By introducing the two hybrid mechanical modes $B_{+}=(G_{1}\delta b_{1}+G_{2}\delta b_{2})/\sqrt{G^{2}_{1}+G^{2}_{2}}$ and $B_{-}=(G_{2}\delta b_{1}-G_{1}\delta b_{2})/\sqrt{G^{2}_{1}+G^{2}_{2}}$, the conditions for the existence of the dark mode in this network-coupled four-mode optomechanical system can be derived as~\cite{Huang2021},
\begin{subequations}
\label{condition}
\begin{align}
M_{1}&=(\omega_{1}-\omega_{2})G_{1}G_{2}+\eta(G_{2}^{2}-G_{1}^{2})=0,\\
M_{2}&=G_{s1}G_{2}-G_{s2}G_{1}=0.\\  \notag
\end{align}
\end{subequations}
Based on these conditions, we can analyze the cases of appearance and disappearance of the dark mode in this system when the two mechanical modes are degenerate ($\omega_{1}=\omega_{2}$). (i) Under the condition of $\eta=0$, i.e., $M_{1}=0$, the hybrid mechanical mode $B_{-}$ decouples from the cavity modes $a$ and $a_{s}$ when $G_{s1}G_{2}-G_{s2}G_{1}=0$ ($G_{s1}/G_{s2}=G_{1}/G_{2}$). In this case, the excitation energy stored in the dark mode $B_{-}$ cannot be extracted via the optomechanical cooling channel, then the thermal noise will destroy the optomechanical entanglement. (ii) When $\eta\neq 0$ and $G_{1}\neq G_{2}$, i.e., $M_{1}\neq0$, we can find that there is no dark mode in this system. (iii) In the case of $\eta\neq 0$ and $G_{1}=G_{2}$, i.e., $M_{1}=0$, $B_{-}$ becomes a dark mode when $G_{s1}=G_{s2}$. However, when $G_{s1}\neq G_{s2}$, i.e., $M_{2}\neq0$, the hybrid mode $B_{-}$ is coupled with the auxiliary cavity mode $a_{s}$, thus the dark mode is broken and optomechanical entanglement can be generated.

To study more clearly the influence of these couplings on the generation of optomechanical entanglement, we consider various coupling structures. Since our intention is to break the dark mode by controlling other four couplings $J$, $G_{s1}$, $G_{s2}$, and $\eta$ [the dashed lines in Fig.~\ref{Fig1}(b)], we keep $G_{1}$ and $G_{2}$ unchanged [the two solid lines in Fig.~\ref{Fig1}(b) always exist] and consider the two-degenerate-mechanical-mode case ($\omega_{1}=\omega_{2}$). In particular, to study the influence of the dark-mode effect on the optomechanical entanglement in this system, we consider the case $G_{1}=G_{2}$ (Note that in the case of $G_{1}\neq G_{2}$, there is a dark mode in the system only when both $\eta=0$ and $G_{1}/G_{2}=G_{s1}/G_{s2}$ are satisfied). Thus, the coupling configurations can be divided into three types: one, two, or three of the four coupling channels $J$, $G_{s1}$, $G_{s2}$, and $\eta$ are switched off~\cite{Huang2021}. For convenience, here we assume $G_{s1}=G_{s2}$ when $G_{s1}$ and $G_{s2}$ exist at the same time. The case $G_{s1}\neq G_{s2}$ will be discussed at the ending of this section. Based on condition~(\ref{condition}), we can find the following results. (i) When one coupling channel is turned off, there are four cases: $J=0$, $\eta=0$, $G_{s1}=0$, and $G_{s2}=0$. When either $J=0$ or $\eta=0$, the dark mode cannot be broken. However, when $G_{s1}=0$ or $G_{s2}=0$, the dark mode can be broken. (ii) When two coupling channels are turned off, there exist six cases: $J=\eta=0$, $G_{s1}=G_{s2}=0$, $J=G_{s1}=0$, $J=G_{s2}=0$, $\eta=G_{s1}=0$, and $\eta=G_{s2}=0$. The dark mode cannot be broken when $J=\eta=0$ or $G_{s1}=G_{s2}=0$. However, in these four cases $J=G_{s1}=0$, $J=G_{s2}=0$, $\eta=G_{s1}=0$, and $\eta=G_{s2}=0$, the dark mode can be broken. (iii) When three coupling channels are turned off, there are four cases: $J=\eta=G_{s1}=0$, $J=\eta=G_{s2}=0$, $J=G_{s1}=G_{s2}=0$, and $\eta=G_{s1}=G_{s2}=0$. The dark mode can (cannot) be broken when $J=\eta=G_{s1}=0$ or $J=\eta=G_{s2}=0$ ($J=G_{s1}=G_{s2}=0$ or $\eta=G_{s1}=G_{s2}=0$).

Corresponding to the above mentioned three types of coupling configurations, we plot in Fig.~\ref{Fig6} the logarithmic negativities $E_{\mathcal{N},1}$ and $E_{\mathcal{N},2}$ as functions of the scaled driving detuning $\Delta_{s}^{\prime\prime}/\omega_{1}$. From Figs.~\ref{Fig6}(a) and~\ref{Fig6}(b) we see  that the optomechanical entanglements are (are not) equal to zero when $J=0$ or $\eta=0$ ($G_{s1}=0$ or $G_{s2}=0$). Figures~\ref{Fig6}(c) and~\ref{Fig6}(d) are plotted in the case where two coupling channels are switched off. We find that the entanglements exist when $J=\eta=0$ or $G_{s1}=G_{s2}=0$, but in these four cases $J=G_{s1}=0$, $J=G_{s2}=0$, $\eta=G_{s1}=0$, and $\eta=G_{s2}=0$, the entanglements disappear. In Figs.~\ref{Fig6}(e) and~\ref{Fig6}(f), the logarithmic negativities $E_{\mathcal{N},1}$ and $E_{\mathcal{N},2}$ are plotted when the three coupling channels are switched off. The results show that the entanglement can exist when $J=\eta=G_{s1}=0$ or $J=\eta=G_{s2}=0$, but cannot exist in the cases of $J=G_{s1}=G_{s2}=0$ or $\eta=G_{s1}=G_{s2}=0$. According to the above results, we find that the performance of the optomechanical entanglement generation depends on whether the dark mode is broken. When there is a dark mode, the entanglement disappears. When the dark mode is broken, a considerable entanglement is created.

\begin{figure}[tbp]
\centering
\includegraphics[width=0.47 \textwidth]{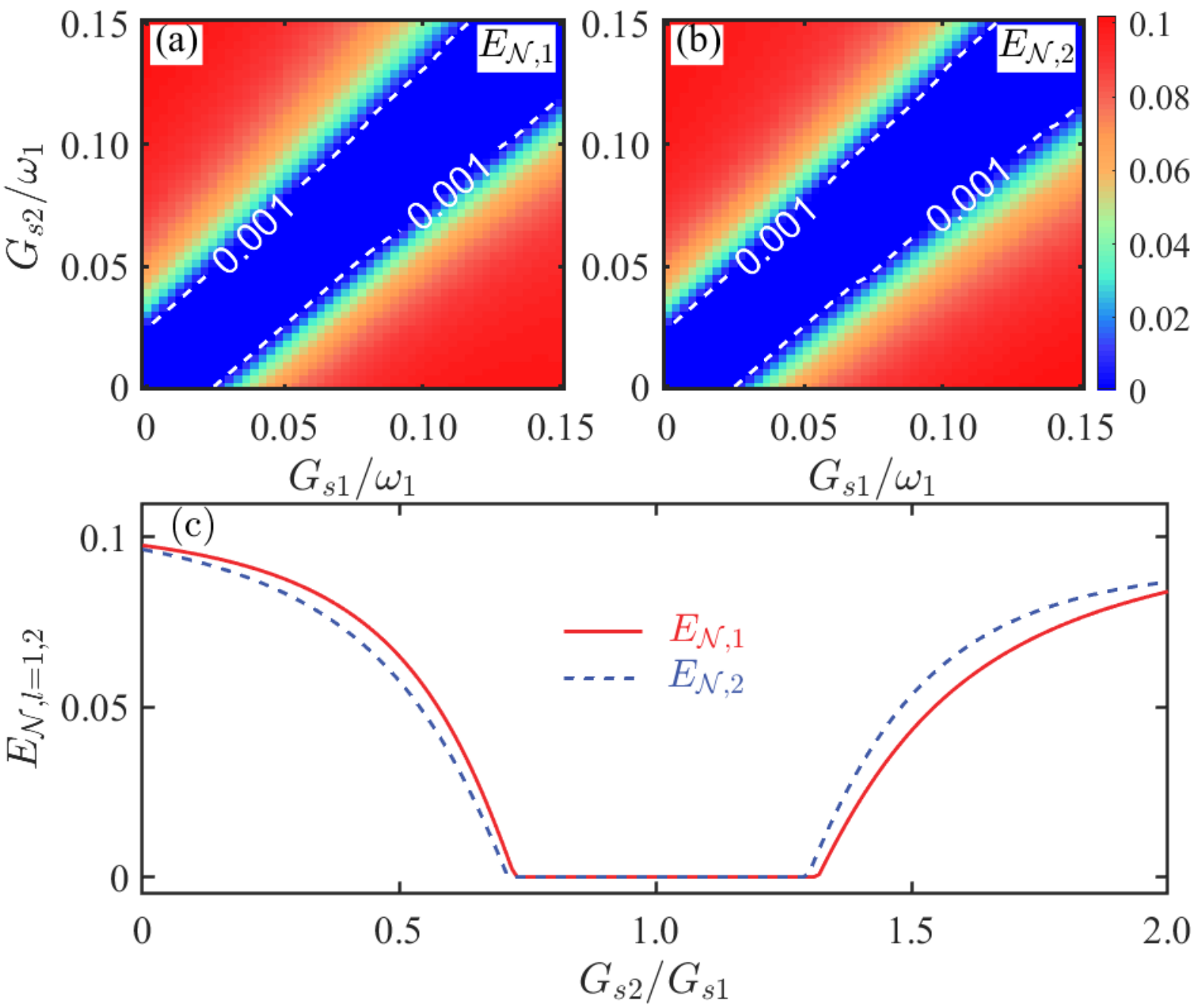}
\caption{(Color online) Logarithmic negativities (a) $E_{\mathcal{N},1}$ and (b) $E_{\mathcal{N},2}$ versus the linearized optomechanical coupling strengths $G_{s1}$ and $G_{s2}$. (c) $E_{\mathcal{N},1}$ and $E_{\mathcal{N},2}$ versus the ratio of the linearized optomechanical coupling strengths $G_{s2}/G_{s1}$. Other parameters are taken as $\omega_{2}/\omega_{1}=1$, $\gamma_{1}/\omega_{1}=\gamma_{2}/\omega_{1}=10^{-5}$, $\kappa/\omega_{1}=\kappa_{s}/\omega_{1}=0.1$, $J/\omega_{1}=\eta/\omega_{1}=0.05$, $\Delta_{c}^{\prime}/\omega_{1}=\Delta_{s}^{\prime\prime}/\omega_{1}=1$, $G_{1}/\omega_{1}=G_{2}/\omega_{1}=0.15$, and $\bar{n}_{1}=\bar{n}_{2}=100$.}
\label{Fig7}
\end{figure}

To better understand the influence of $G_{s1}$ and $G_{s2}$ on the generation of optomechanical entanglement, we plot in Figs.~\ref{Fig7}(a) and~\ref{Fig7}(b) the logarithmic negativities $E_{\mathcal{N},1}$ and $E_{\mathcal{N},2}$ as functions of the scaled coupling strengths $G_{s1}/\omega_{1}$ and $G_{s2}/\omega_{1}$. Here we see that the optomechanical entanglements disappear when $G_{s1}$ and $G_{s2}$ have equal or close values, as shown by the valley area along the diagonal line $G_{s1}=G_{s2}$. This feature can be seen more clearly in Figs.~\ref{Fig7}(c), $E_{\mathcal{N},1}$ and $E_{\mathcal{N},2}$ are zero during the intermediate interval of $G_{s2}/G_{s1}=0.7$-$1.3$. These results are consistent with the analyses obtained based on Eq.~(\ref{condition}). For finite values of the ratio $G_{s2}/G_{s1}$, the dark-mode effect works, then the generation of optomechanical entanglements between cavity mode and two mechanical modes is unfeasible. When $G_{s2}/G_{s1}<0.5$ or $G_{s2}/G_{s1}>1.5$, the dark-mode effect is broken, then a large entanglement can be obtained.

\section{Discussions on the experimental implementation and the dark-mode-breaking mechanism for the nonlinear optomechanical coupling  \label{Discuss}}

In this section  we present discussion of the experimental implementation of our scheme and elaborate the physical mechanism of the dark-mode breaking in the absence of the optomechanical-coupling linearization.

\subsection{Discussions on the experimental implementation\label{DiscussA}}

\begin{figure}[tbp]
\centering
\includegraphics[width=0.47 \textwidth]{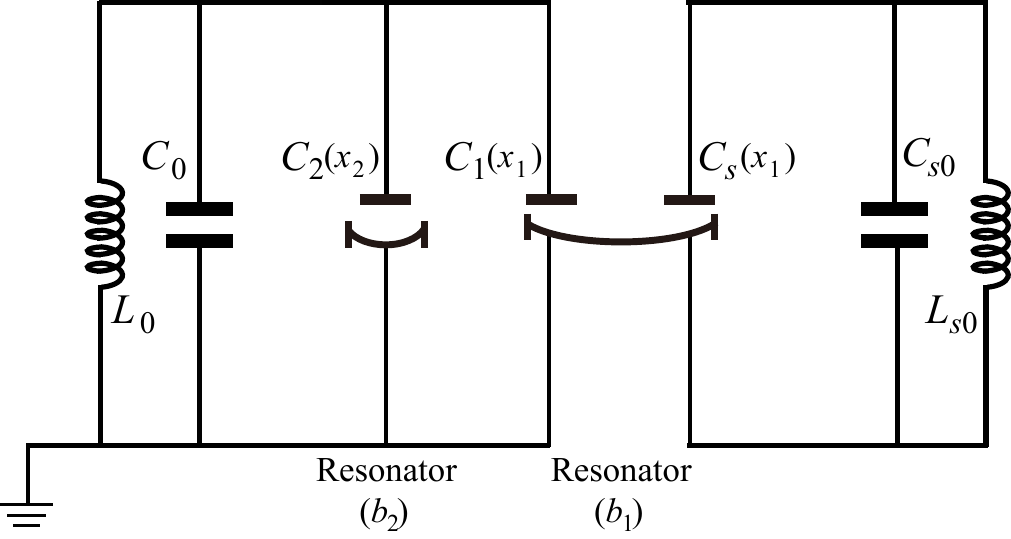}
\caption{Schematic of a microwave electromechanical system. The microwave cavity on the left consists of an inductor (with inductance $L_{0}$) and three capacitors [with capacitances $C_{0}$, $C_{1}(x_{1})$, and $C_{2}(x_{2})$]. The microwave cavity on the right consists of an inductor (with inductance $L_{s0}$) and two capacitors [with capacitances $C_{s0}$ and $C_{s}(x_{1})$]. Here the two capacitances $C_{1}(x_{1})$ and $C_{s}(x_{1})$ depend on a common micromechanical resonator $b_{1}$ and the capacitance $C_{2}(x_{2})$ depends on the micromechanical resonator $b_{2}$. The displacements $x_{l=1,2}$ of the mechanical resonators modulate the capacitances and hence the frequencies of microwave cavities. Note that the phonon-hopping coupling between the two mechanical resonators can be realized by the stress manipulation and the photon-hopping coupling between the two microwave cavities can be induced by a capacitor.}
\label{Fig8}
\end{figure}

We now discuss the experimental implementation of this auxiliary-cavity-mode scheme with microwave optomechanical systems~\cite{Teufel2011a,Ockeloen-Korppi2018,Lepinay2021,Kotler2021,Ockeloen-Korppi2016a,Ockeloen-Korppi2016b}. Concretely, we consider an inductor-capacitor (LC) oscillator consisting of an inductor (with inductance $L_{0}$) and three capacitors [with capacitances $C_{0}$, $C_{1}(x_{1})$, and $C_{2}(x_{2})$], as shown in Fig.~\ref{Fig8}. The two capacitances $C_{1}(x_{1})$ and $C_{2}(x_{2})$ depend on the micromechanical resonators $b_{1}$ and $b_{2}$, respectively. To achieve the auxiliary-cavity-mode scheme, we introduce an auxiliary LC oscillator, which consists of an inductor (with inductance $L_{s0}$) and two capacitors [with capacitances $C_{s0}$ and $C_{s}(x_{1})$]. Note that $C_{1}(x_{1})$ and $C_{s}(x_{1})$ are mounted on a common micromechanical resonator $b_{1}$. Moreover, there is no electric circuit connection between the two LC oscillators for realizing the Hamiltonian in Eq.~(\ref{Hamit1}). However, for the introducing photon-hopping coupling between the two cavity modes $a$ and $a_{s}$, a capacitor is needed to couple the two LC resonators.

To better understand the experimental circuit diagram, below we present a detailed derivation of the effective Hamiltonian of this microwave optomechanical system. For the superconducting circuit on the left, the total kinetic energy $T$ is given by
\begin{equation}
T=\frac{1}{2}C_{0}\dot{\Phi}^{2}+\frac{1}{2}C_{1}\left( x_{1}\right) \dot{
\Phi}^{2}+\frac{1}{2}C_{2}\left( x_{2}\right) \dot{\Phi}^{2},
\end{equation}
where $\Phi$ is the generalized magnetic flux. The energy of the inductor is identified as the potential energy $U=\Phi ^{2}/2L_{0}$. The Lagrangian of this circuit can be written as
\begin{eqnarray}
\mathcal{L} &=&T-U \nonumber\\
&=&\frac{1}{2}\left[ C_{0}+C_{1}\left( x_{1}\right) +C_{2}\left(
x_{2}\right) \right] \dot{\Phi}^{2}-\frac{1}{2L_{0}}\Phi^{2}.
\end{eqnarray}
By introducing the momentum conjugate to the generalized magnetic flux $\Phi$
\begin{equation}
P=\frac{\partial \mathcal{L}}{\partial \dot{\Phi}}=[C_{0}
+C_{1}\left( x_{1}\right) +C_{2}\left( x_{2}\right)]\dot{\Phi},
\end{equation}
the Hamiltonian of this circuit can be derived as
\begin{eqnarray}
H_{L}&=&\frac{\partial \mathcal{L}}{\partial \dot{\Phi}}\dot{\Phi}-\mathcal{L}\nonumber\\
&=&\frac{P^{2}}{2C(x_{1},x_{2})}+\frac{1}{2L_{0}}\Phi^{2},
\end{eqnarray}
where we introduce the variable $C(x_{1},x_{2})=C_{0}+C_{1}( x_{1})+C_{2}(x_{2})$. Since the mechanical displacements in $C(x_{1},x_{2})$ are small, we can expand the variable $C(x_{1},x_{2})$ to the first order of $x_{1}$ and $x_{2}$, then the Hamiltonian can be expressed as
\begin{eqnarray}
H_{L}&\approx&\frac{P^{2}}{2C\left(0,0\right)}+\frac{1}{2L_{0}}\Phi ^{2}-\frac{P^{2}}{2C^{2}(0,0)}\nonumber\\
&&\times\left[ \left.\frac{\partial C\left(
x_{1},x_{2}\right) }{\partial x_{1}}\right\vert_{x_{1}=x_{2}=0} x_{1}+
\left.\frac{\partial C\left( x_{1},x_{2}\right) }{\partial x_{2}}\right\vert
_{x_{1}=x_{2}=0}x_{2}\right].\nonumber\\ \label{Haee}
\end{eqnarray}

Using the canonical quantization, the momentum and flux operators can be expressed with the creation and annihilation operators as
\begin{eqnarray}
P&=&i\sqrt{\frac{\hbar}{2L_{0}\omega_{0}}}(a^{\dag}-a), \nonumber\\
\Phi &=&\sqrt{\frac{\hbar L_{0}\omega_{0}}{2}}(a+a^{\dag}), \label{pphi}
\end{eqnarray}
which obey the commutation relation $[\Phi,P]=i\hbar$. Here, we introduce the oscillation frequency $\omega_{0}=1/\sqrt{L_{0}C(0,0)}$. Denote the modulation frequency as $\omega({x_{1},x_{2}})=1/\sqrt{L_{0}C(x_{1},x_{2})}$, then we obtain the relationship
\begin{eqnarray}
\left. \frac{\partial C\left( x_{1},x_{2}\right) }{\partial x_{1}}\right\vert
_{x_{1}=x_{2}=0} &=&-\frac{2\eta_{1}}{L_{0}\omega _{0}^{3}},\nonumber\\
\left. \frac{\partial C\left( x_{1},x_{2}\right) }{\partial x_{2}}\right\vert
_{x_{1}=x_{2}=0} &=&-\frac{2\eta_{2}}{L_{0}\omega _{0}^{3}},\label{cx12}
\end{eqnarray}
where we introduce the parameters $\eta_{1}=[\partial \omega( x_{1},x_{2})/\partial x_{1}]_{x_{1}=x_{2}=0}$ and $\eta_{2}=[\partial\omega(x_{1},x_{2})/\partial x_{2}] _{x_{1}=x_{2}=0}$. Substituting Eqs. ~(\ref{pphi}) and ~(\ref{cx12}) into Hamiltonian~(\ref{Haee}), we obtain ($\hbar=1$)
\begin{eqnarray}
H_{L}&\approx&\omega_{0}a^{\dag }a-\frac{1}{2}(a^{\dag }a^{\dag}+aa-2a^{\dag
}a-1)(\eta_{1}x_{1}+\eta_{2}x_{2}) \nonumber \\
&\approx &\omega_{0}a^{\dag}a+\frac{1}{2}( 2a^{\dag}a+1)(\eta_{1}x_{1}+\eta_{2}x_{2}).\label{Haeee2}
\end{eqnarray}
In the second line of Eq.~(\ref{Haeee2}), we neglected the two high-frequency oscillating terms ($a^{\dag}a^{\dag}$ and $aa$) under the condition that the resonant frequency of the optical mode is much larger than the resonant frequencies $\omega_{1}$ and $\omega_{2}$ of the mechanical modes.

The displacements $x_{l=1,2}$ can be quantized using the phonon creation and annihilation operators $x_{l=1,2}=x_{l,\text{zpf}}(b_{l}^{\dag}+b_{l})$, with $x_{l,\text{zpf}}$ being zero-point fluctuation. Then the Hamiltonian~(\ref{Haeee2}) is reduced to
\begin{eqnarray}
H_{L}&\approx&\omega _{0}a^{\dag }a+\sum_{l=1,2}[\omega_{l}b_{l}^{\dagger}b_{l}+g^{\prime}_{l}a^{\dagger}a(b_{l}^{\dagger}+b_{l})]\nonumber\\
&&+\sum_{l=1,2}\frac{1}{2}[g^{\prime}_{l}( b_{l}^{\dag}+b_{l})],
\end{eqnarray}
where we introduce the coupling strengths $g^{\prime}_{l=1,2}=\eta_{l}x_{l,\text{zpf}}$. For the Hamiltonian $H_{L}$, we introduce
the displacement operators $D_{l}(\beta_{l})=\exp[\beta_{l}(b_{l}^{\dag}-b_{l})]$ for $l=1,2$, with the parameters $\beta_{l=1,2}=-g^{\prime}_{l}/2\omega_{l}$. Then the transformed Hamiltonian becomes
\begin{eqnarray}
\tilde{H}_{L}&=&D^{\dagger}_{1}(\beta_{1})D^{\dagger}_{2}(\beta_{2})H_{L}D_{1}(\beta_{1})D_{2}(\beta_{2})\nonumber\\
&&\approx\omega_{c}a^{\dag}a+\sum_{l=1,2}[\omega_{l}b_{l}^{\dagger}b_{l}+g^{\prime}_{l}a^{\dagger}a(b_{l}^{\dagger}+b_{l})],
\end{eqnarray}
where $\omega_{c}=\omega_{0}-g^{\prime2}_{1}/\omega_{l}-g^{\prime2}_{2}/\omega_{2}$ is the effective frequency of the cavity mode.

By adopting the same method, we can obtain the Hamiltonian of the superconducting circuit on the right as
\begin{equation}
H_{R}\approx\omega_{s}a_{s}^{\dag }a_{s}+ g^{\prime}_{s1}a_{s}^{\dag }a_{s}(b_{1}^{\dag }+b_{1}).\\
\end{equation}
Based on the above discussions, we know that the total Hamiltonian of the microwave optomechanical system can be written as
\begin{eqnarray}
H_{T}&\approx&\omega_{c}a^{\dag}a+\sum_{l=1,2}[\omega_{l}b_{l}^{\dagger}b_{l}+g^{\prime}_{l}a^{\dagger}a(b_{l}^{\dagger}+b_{l})]\nonumber \\
&&+\omega_{s}a_{s}^{\dag }a_{s}+ g^{\prime}_{s1}a_{s}^{\dag }a_{s}(b_{1}^{\dag }+b_{1}).
\end{eqnarray}
The above Hamiltonian shows that our auxiliary-cavity-mode scheme can be experimentally achieved in the superconducting circuit. In particular, in this setup the phonon-hopping interaction between the two mechanical resonators can be introduced either by coupling them to a superconducting charge qubit~\cite{Lai2020PRARC,Lai2022} or through the stress manipulation~\cite{Okamoto2013,Huang2013,Huang2016}, and the photon-hopping interaction between the two microwave cavities can be induced by a capacitor~\cite{Yamamoto2014,Peterson2017}.

Based on the current experimental progress, we present some estimations on the experimental parameters of the microwave optomechanical systems~\cite{Teufel2011a,Ockeloen-Korppi2018,Lepinay2021,Kotler2021,Ockeloen-Korppi2016a,Ockeloen-Korppi2016b}. As shown in Fig.~\ref{Fig8}, the two effective microwave cavities are two LC oscillators with resonance frequencies $\omega_{c}=\omega_{s}\approx2\pi\times7$ GHz and decay rates $\kappa=\kappa_{s}\approx2\pi\times1$ MHz~\cite{Ockeloen-Korppi2016a,Ockeloen-Korppi2016b}. The left microwave cavity acts as the intermediate cavity mode $a$, and the right microwave cavity acts as the auxiliary cavity mode $a_{s}$. The two micromechanical resonators are fabricated on quartz substrates, with resonance frequencies $\omega_{1}= \omega_{2}\approx2\pi\times 8$ MHz and decay rates $\gamma_{1}=\gamma_{2}\approx 2\pi\times100$ Hz. The motion of the micromechanical resonators modulates the resonance frequencies of two microwave cavities, thus realizing the optomechanical interactions. The coupling strengths between the microwave cavity $a$ and two mechanical resonators are taken as $G_{1}=G_{2}\approx2\pi\times 0.5$ MHz, and the coupling strengths between the microwave cavity $a_{s}$ and two mechanical resonators are $G_{s1}=G_{s2}\approx 2\pi\times0.8$ MHz.

According to the above experimental parameters, the scaled parameters in our simulations are taken as $\omega_{2}=\omega_{1}$, $\gamma_{1}/\omega_{1}=\gamma_{2}/\omega_{1}=10^{-5}$, $G_{1}/\omega_{1}=G_{2}/\omega_{1}=0.15$, $G_{s1}/\omega_{1}=G_{s2}/\omega_{1}=0\text{-}0.15$, $J/\omega_{1}=\eta/\omega_{1}=0.05$, $\Delta_{c}^{\prime}/\omega_{1}=\Delta_{s}^{\prime}/\omega_{1}=1$, $\kappa/\omega_{1}=\kappa_{s}/\omega_{1}=0.1$, and $\bar{n}_{1}=\bar{n}_{2}=100$. These used parameters are of the same order of the experimental parameters. These analyses indicate that our scheme is feasible under the present experimental conditions.

\subsection{Discussion on the dark-mode-breaking mechanism in the absence of the optomechanical-coupling linearization}

\begin{figure}[tbp]
\centering
\includegraphics[width=0.47 \textwidth]{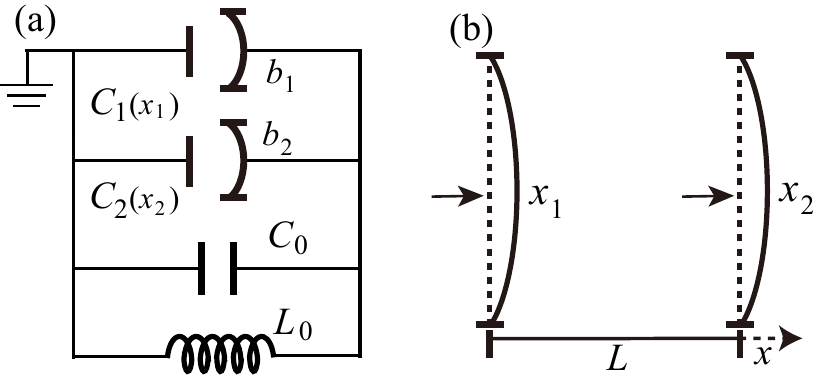}
\caption{(a) The microwave electromechanical system consists of an inductor (with inductance $L_{0}$) and three capacitors [with capacitances $C_{0}$, $C_{1}(x_{1})$, and $C_{2}(x_{2})$]. Here the two capacitances $C_{1}(x_{1})$ and $C_{2}(x_{2})$ depend on two  micromechanical resonators $b_{1}$ and $b_{2}$, respectively. The displacements $x_{l=1,2}$ of the mechanical resonators modulate the resonance frequency of the microwave cavity. (b) A Fabry-Perot-type optomechanical cavity (length $L$) with two movable mirrors.   The radiation-pressure force exerted by the photons causes the motion of two mirrors, and then the displacements $x_{l=1,2}$ of two mirrors modulate the resonance frequency of the optical cavity.}
\label{Fig9}
\end{figure}

In our above discussions, we analyzed the dark-mode effect by introducing new hybrid mechanical modes after the linearization of the optomechanical interaction. Based on the coupling configuration in the four-mode optomechanical system, it is natural to ask what is the relationship between the dark-mode effect and the center-of-mass and relative coordinates. In this section, we analyze the physical mechanism behind the dark-mode breaking in the absence of optomechanical-coupling linearization. The Hamiltonian of the four-mode optomechanical system derived from Fig.~\ref{Fig8} can be rewritten as
\begin{eqnarray}
H_{P} &=&\Delta _{c}a^{\dag }a+\Delta _{s}a_{s}^{\dag }a_{s}+\sum_{l=1,2}
\left[ \frac{\omega _{l}}{2}\left(p_{l}^{2}+q_{l}^{2}\right) +\tilde{g}_{l}a^{\dag }aq_{l}\right]  \nonumber\\
&&+\tilde{g}_{s1}a_{s}^{\dag }a_{s}q_{1}+( \Omega a^{\dag }+\Omega_{s}a_{s}^{\dag }+\text{H.c.}),
\label{Hamit1p}
\end{eqnarray}
where $p_{l}$ and $q_{l}$ are the dimensionless momentum and position operators of the $l$th mechanical mode, respectively. The $\tilde{g}_{l=1,2}$ ($\tilde{g}_{s1}$) term describes the  optomechanical coupling between the cavity mode $a$ $(a_{s})$ and the $l$th mechanical mode $b_{l}$ ($b_{1}$). Note that the relationships between the coupling parameters defined in Eq.~(\ref{Hamit1p}) and those presented in Eq.~(\ref{Haeee2}) are given by $\tilde{g}_{l=1,2}=\eta_{l=1,2}$ and $\tilde{g}_{s1}=\eta_{s1}$ with $\eta_{s1}=[\partial \omega( x_{1})/\partial x_{1}]_{x_{1}=0}$. The other variables have been defined in Eq.~(\ref{Hamit1}).

To observe the dark-mode effect in the system when the two mechanical modes are degenerate, we first consider the case where the auxiliary cavity mode does not exist, then the Hamiltonian reads
\begin{equation}
\tilde{H}_{P}=\Delta _{c}a^{\dag }a+\sum_{l=1,2}\left[ \frac{\omega _{l}}{2}\left( p_{l}^{2}+q_{l}^{2}\right) +\tilde{g}_{l}a^{\dag }aq_{l}\right]
+\Omega(a^{\dag}+a). \label{Hamiltutetp}\\
\end{equation}
To describe the dark-mode effect, we introduce the effective center-of-mass and relative coordinates
\begin{eqnarray}
q_{\text{cm}} &=&\frac{\tilde{g}_{1}q_{1}+\tilde{g}_{2}q_{2}}{\sqrt{\tilde{g}
_{1}^{2}+\tilde{g}_{2}^{2}}}, \hspace{0.5 cm}  p_{\text{cm}}=\frac{\tilde{g}_{1}p_{1}+\tilde{g}
_{2}p_{2}}{\sqrt{\tilde{g}_{1}^{2}+\tilde{g}_{2}^{2}}}, \nonumber\\
q_{\text{r}} &=&\frac{\tilde{g}_{1}q_{2}-\tilde{g}_{2}q_{1}}{\sqrt{\tilde{g}
_{1}^{2}+\tilde{g}_{2}^{2}}}, \hspace{0.5 cm} p_{\text{r}}=\frac{\tilde{g}_{1}p_{2}-\tilde{g}
_{2}p_{1}}{\sqrt{\tilde{g}_{1}^{2}+\tilde{g}_{2}^{2}}}.
\end{eqnarray}
Then we can reexpress the momentum and position operators with the center-of-mass and relative degrees of freedom as
\begin{eqnarray}
q_{1} &=&\frac{\tilde{g}_{1}q_{\text{cm}}-\tilde{g}_{2}q_{\text{r}}}{\sqrt{\tilde{g}_{1}^{2}+\tilde{g}_{2}^{2}}}, \hspace{0.5 cm} q_{2}=\frac{\tilde{g}_{2}q_{\text{cm}}+\tilde{g}_{1}q_{\text{r}}}{\sqrt{\tilde{g}_{1}^{2}+\tilde{g}_{2}^{2}}}, \nonumber\\
p_{2} &=&\frac{\tilde{g}_{2}p_{\text{cm}}+\tilde{g}_{1}p_{\text{r}}}{\sqrt{\tilde{g}_{1}^{2}+\tilde{g}_{2}^{2}}}, \hspace{0.5 cm}   p_{1}=\frac{\tilde{g}_{1}p_{\text{cm}}-\tilde{g}_{2}p_{\text{r}}}{\sqrt{\tilde{g}_{1}^{2}+\tilde{g}_{2}^{2}}}.
\end{eqnarray}
With these new coordinates, the Hamiltonian $\tilde{H}_{P}$ in Eq.~(\ref{Hamiltutetp}) becomes
\begin{eqnarray}
\tilde{H}_{P} &=&\Delta _{c}a^{\dag }a+\frac{\omega _{\text{cm}}}{2}(p_{\text{cm}}^{2}+q_{\text{cm}}^{2}) +\frac{\omega _{\text{r}}}{2}
(p_{\text{r}}^{2}+q_{\text{r}}^{2})+\Omega(a^{\dag }+a) \nonumber\\
&&+\frac{(\omega_{2}-\omega_{1})\tilde{g}_{1}\tilde{g}_{2}}{\tilde{g}_{1}^{2}+\tilde{
g}_{2}^{2}}(p_{\text{cm}}p_{\text{r}}+q_{\text{cm}}q_{\text{r}})-\sqrt{\tilde{g}_{1}^{2}+\tilde{g}_{2}^{2}}a^{\dag }aq_{\text{cm}},\notag\\
\label{Hamilttp}
\end{eqnarray}
where we introduce the effective frequencies
\begin{subequations}
\begin{align}
\omega_{\text{cm}}=&\frac{\omega_{1}\tilde{g}_{1}^{2}+\omega_{2}\tilde{g}_{2}^{2}}{\tilde{g}_{1}^{2}+\tilde{g}_{2}^{2}}, \\
\omega_{\text{r}}=&\frac{\omega_{1}\tilde{g}_{2}^{2}+\omega_{2}\tilde{g}_{1}^{2}}{\tilde{g}_{1}^{2}+\tilde{g}_{2}^{2}}.
\end{align}
\end{subequations}
Equation~(\ref{Hamilttp}) shows that in the degenerate-resonator case ($\omega _{1}=\omega _{2}$), the relative coordinate $q_{\text{r}}$ is decoupled from both the center-of-mass coordinate $q_{\text{cm}}$ and the cavity mode $a$. As a consequence, the relative coordinate $q_{\text{r}}$ becomes a dark mode.

To break the dark mode, we present the auxiliary-cavity-mode method by introducing an auxiliary cavity (with frequency $\omega_{s}$) coupled with the mechanical mode $q_{1}$ and a driving to the auxiliary cavity (with amplitude $\Omega_{s}$ and frequency $\omega_{d}=\omega_{s}-\Delta_{s}$), then the Hamiltonian in Eq.~(\ref{Hamit1p}) can be written as
\begin{eqnarray}
H_{P} &=&\Delta _{c}a^{\dag }a+\Delta _{s}a_{s}^{\dag }a_{s}+\frac{\omega_{\text{cm}}}{2}(p_{\text{cm}}^{2}+q_{\text{cm}}^{2})+\frac{
\omega _{\text{r}}}{2}(p_{\text{r}}^{2}+q_{\text{r}}^{2}) \nonumber\\
&&+\frac{(\omega_{2}-\omega _{1})\tilde{g}_{1}\tilde{g}_{2}}{\tilde{g}_{1}^{2}+\tilde{g}_{2}^{2}}(p_{\text{cm}}p_{\text{r}}+q_{\text{cm}}q_{\text{r}}) -\sqrt{\tilde{g}_{1}^{2}+\tilde{g}_{2}^{2}}a^{\dag }aq_{\text{cm}} \nonumber\\
&&+\tilde{g}_{s1}a_{s}^{\dag }a_{s}\frac{\tilde{g}_{1}q_{\text{cm}}-\tilde{g}
_{2}q_{\text{r}}}{\sqrt{\tilde{g}_{1}^{2}+\tilde{g}_{2}^{2}}}+(\Omega a^{\dag}+\Omega_{s}a_{s}^{\dag }+\text{H.c.}).\label{Hamipq}
\end{eqnarray}
We can see from $H_{P}$ in Eq.~(\ref{Hamipq}) that the relative coordinate $q_{\text{r}}$ is always coupled with the auxiliary cavity mode $a_{s}$. Therefore, even if the relative coordinate $q_{\text{r}}$ is decoupled from both the center-of-mass coordinate $q_{\text{cm}}$ and the cavity mode $a$, the dark mode can also be broken via the auxiliary cavity mode $a_{s}$.

We want to emphasize that the relation between the dark/bright mode and the relative/center-of-mass coordinate depends on the realistic physical system. This is because the form of the relative/center-of-mass coordinate depends on the choice of the common coordinate system. Below, we compare the form of the radiation-pressure interactions in the microwave electromechanical system [Fig.~\ref{Fig9}(a)] and Fabry-Perot-type cavity optomechanical system [Fig.~\ref{Fig9}(b)]. For the microwave electromechanical system, when $\tilde{g}_{l}=\tilde{g}_{2}=\tilde{g}$, the radiation-pressure interactions between the microwave cavity and two mechanical modes can be written as (see Sec.~\ref{DiscussA} for detailed derivation)
\begin{eqnarray}
H_{I} &=& \tilde{g}a^{\dag }a(x_{1}+x_{2}).
\label{Hamitinter}
\end{eqnarray}
In this case, the center-of-mass coordinate  $q_{\text{cm}}$ becomes a bright mode, which is coupled to the cavity mode. The relative coordinate $q_{\text{r}}$ is decoupled from both the center-of-mass coordinate $q_{\text{cm}}$ and the cavity mode $a$, and it becomes a dark mode.

In the Fabry-Perot-type cavity optomechanical system, the resonance frequency of the cavity field is
 \begin{eqnarray}
\omega_{c}(x_{1},x_{2}) &=&\frac{n\pi c}{L+x_{2}-x_{1}}\approx\frac{n\pi c}{L}\left(1-\frac{x_{2}-x_{1}}{L}\right)  \nonumber\\
&=&\omega_{c}\left(1-\frac{x_{2}-x_{1}}{L}\right),
\label{Hamitinterp}
\end{eqnarray}
where the parameters $c$, $L$, and $x_{l=1,2}$ are the speed of light in vacuum, the length of the rest cavity, and the displacements of the two mirrors, respectively. In this case, the optomechanical couplings between the cavity mode and two mechanical modes can be expressed as
 \begin{eqnarray}
H^{\prime}_{I}&=&ga^{\dag }a(x_{1}-x_{2}),
\label{Hamitinterp}
\end{eqnarray}
where  $g=g_{l=1,2}=\omega_{c}x_{l,\text{zpf}}/L$ is the single-photon optomechanical-coupling strength. Here, the center-of-mass coordinate  $q_{\text{cm}}$ is decoupled from the cavity mode and becomes a dark mode. But the relative coordinate $q_{\text{r}}$ is coupled from the cavity mode $a$, and it becomes a bright mode.

\subsection{Clarification the dominate role of the dark-mode breaking in the enhancement of  optomechanical entanglement}

\begin{figure}[tbp]
\centering
\includegraphics[width=0.4 \textwidth]{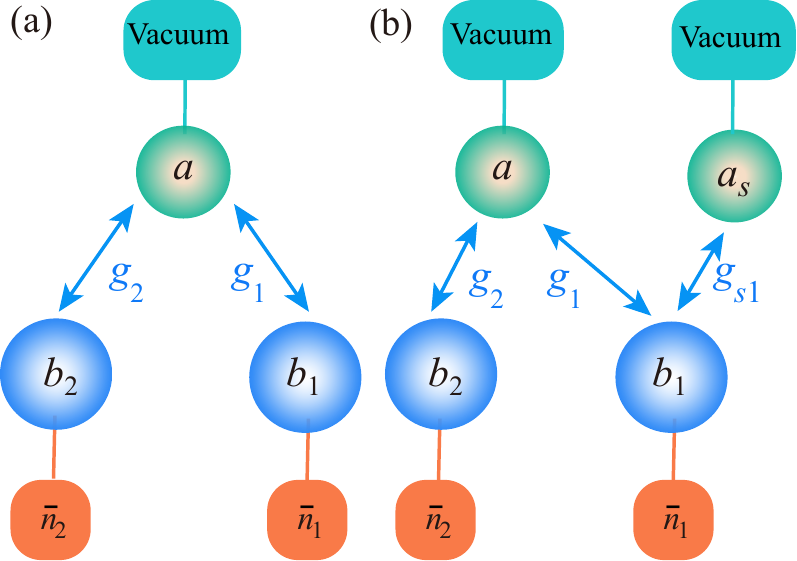}
\caption{(a) Schematic of the three-mode optomechanical system consisting of a cavity mode and two mechanical modes. The cavity mode $a$ is coupled to the $l$th mechanical mode $b_{l=1,2}$ via the optomechanical interaction with the coupling strength $g_{l}$. The cavity mode $a$ and the $l$th mechanical mode $b_{l}$ are coupled to the vacuum bath and the heat bath (thermal-occupation number $\bar{n}_{l}$), respectively. (b) An auxiliary cavity mode $a_{s}$ coupled with a vacuum bath is introduced into the three-mode optomechanical system, and it is coupled to the mechanical mode $b_{1}$ via the optomechanical interaction of strength $g_{s1}$.}
\label{Fig10}
\end{figure}

In Sec.~\ref{entanglement}, we have studied the generation of optomechanical entanglement between the cavity mode and two mechanical modes by breaking the dark-mode effect with the auxiliary cavity mode, when the mechanical baths are at nonzero temperatures. The introduction of the auxiliary cavity mode also brings  an additional degree of freedom. Thus a natural question arising is how does the correlation established between the auxiliary cavity mode $a_{s}$ and the mechanical mode $b_{1}$ affect the optomechanical entanglement between the cavity mode $a$ and the two mechanical modes $b_{1}$ and $b_{2}$. We know that both the introduced correlation and the dark-mode breaking will affect the optomechanical entanglement. However, which one is the dominate factor remains unclear. To clarify this point, we now study optomechanical entanglement between the cavity mode and the two mechanical modes in both the dark-mode-unbreaking and -breaking cases when the mechanical baths are at either zero or finite ($\bar{n}=100$) temperatures.

In Fig.~\ref{Fig10} we plot the simplified schematic of the three-mode and four-mode optomechanical systems. As shown in Fig.~\ref{Fig10}(a), the cavity mode $a$ is optomechanically coupled to two mechanical modes $b_{l=1,2}$ of strengths $g_{l=1,2}$. Moreover, the cavity mode $a$ and the $l$th mechanical mode $b_{l}$ are coupled to the individual vacuum bath  and the individual heat bath (thermal-occupation number $\bar{n}_{l}$), respectively. In the degenerate-mechanical-mode case, one of the two mechanical normal modes becomes the dark mode. To break the mechanical dark mode, we introduce an auxiliary cavity mode $a_{s}$ to couple with the mechanical mode $b_{1}$ via the optomechanical interaction of strength $g_{s1}$, and the auxiliary cavity mode $a_{s}$ is coupled to the individual vacuum bath  [see Fig.~\ref{Fig10}(b)].

\begin{figure}[tbp]
\centering
\includegraphics[width=0.47 \textwidth]{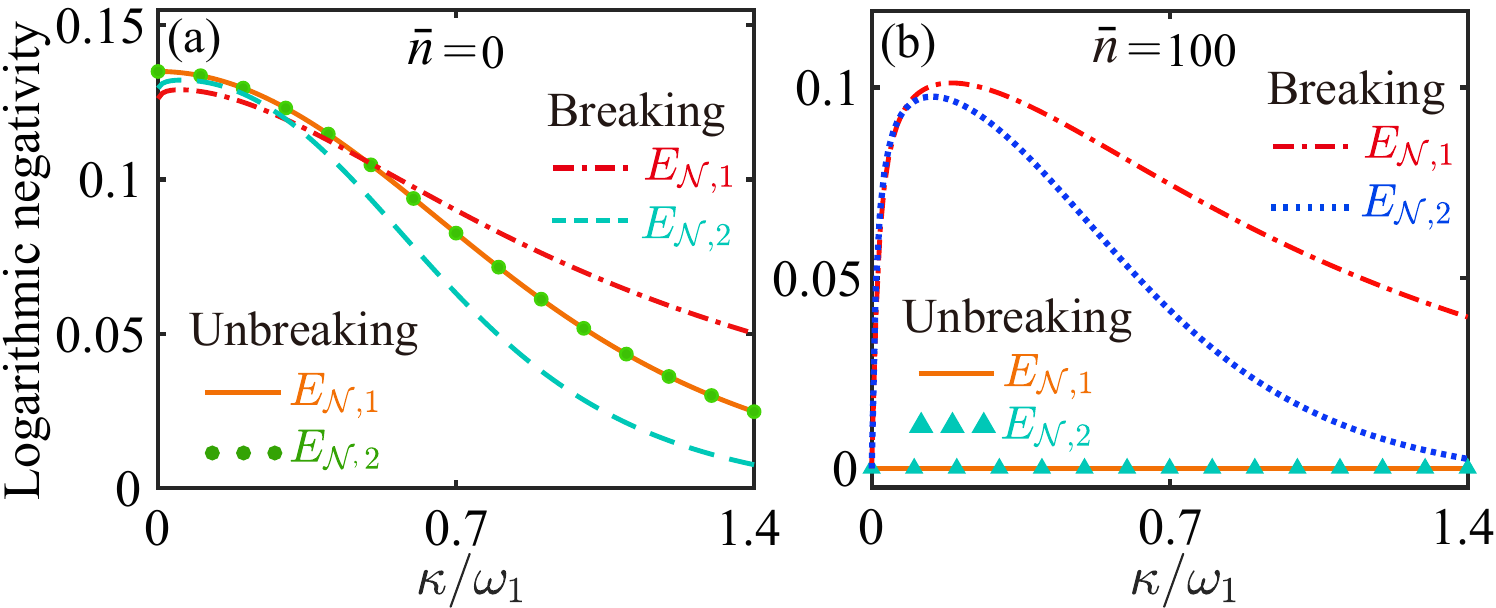}
\caption{Logarithmic negativities $E_{\mathcal{N},1}$ and $E_{\mathcal{N},2}$ versus the scaled decay rate $\kappa/\omega_{1}$
  in both the dark-mode-unbreaking  and -breaking cases when the thermal phonon numbers (a) $\bar{n}=0$ and (b) $\bar{n}=100$. Other parameters used are $\omega_{2}=\omega_{1}$, $\bar{n}_{1}=\bar{n}_{2}=\bar{n}$, $\gamma_{1}/\omega_{1}=\gamma_{2}/\omega_{1}=10^{-5}$, $G_{1}/\omega_{1}=G_{2}/\omega_{1}=0.15$, $G_{s1}/\omega_{1}=0.1$, $\Delta_{c}^{\prime}/\omega_{1}=\Delta_{s}^{\prime}/\omega_{1}=1$, and $\kappa_{s}/\omega_{1}=0.1$.}
\label{Fig11}
\end{figure}

In Fig.~\ref{Fig11} we plot the logarithmic negativities $E_{\mathcal{N},1}$ and $E_{\mathcal{N},2}$ as functions of the scaled decay rate $\kappa/\omega_{1}$ in both the dark-mode-unbreaking and -breaking cases when the mechanical baths are at either zero or nonzero temperatures. As shown in  Fig.~\ref{Fig11} (a), the logarithmic negativities $E_{\mathcal{N},1}$ and $E_{\mathcal{N},2}$ decrease with the increase of the scaled decay rate $\kappa/\omega_{1}$ in both the dark-mode-unbreaking and -breaking cases when the mechanical baths are at zero temperatures. Here, we choose $\bar{n}=0$ such that the crosstalk from the thermal noise can be avoided and that we can analyze the pure influence of the correlation between $a_{s}$ and  $b_{1}$ on the optomechanical entanglement. We observe that, in the absence of mechanical thermal noise ($\bar{n}=0$), the influence of the auxiliary cavity mode on the optomechanical entanglement is weak. In the presence of the thermal noise (for example $\bar{n}=100$), differently, we find that the breaking of the dark mode will largely increase the optomechanical entanglement, as  shown in Fig.~\ref{Fig11}(b). This indicates that the dark-mode effect is the dominate factor for affecting the generation of optomechanical entanglement.

The physical reason behind those phenomena can be understood as follows. In the absence of the auxiliary cavity mode, the dark mode decouples from the cavity mode and prevents the energy extraction through the optomechanical cooling channel, then the thermal excitations stored in the dark mode will destroy the optomechanical entanglement. In this case, the thermal noise in the dark mode is the dominated negative factor for the entanglement. By introducing the auxiliary cavity mode, a new cooling channel will be built to extract the thermal excitations stored in the dark mode. As a result, the ground-state cooling can be realized and the optomechanical entanglement can be created by breaking the dark mode. Though the auxiliary cavity will create the mixing of the state of the cavity and mechanical modes, the thermal noise is the vital factor for degrading the optomechanical entanglement.

\section{Conclusion\label{conclu}}

In conclusion, we have studied the generation of optomechanical entanglement in a four-mode optomechanical system consisting of two degenerate or near-degenerate mechanical modes and two cavity modes. Here, the common coupling cavity mode is coupled to two mechanical modes, while the auxiliary cavity mode is coupled to one of the two mechanical modes. The motivation for introducing the auxiliary cavity mode is to break the dark mode formed by the two degenerate mechanical modes coupled to a common cavity mode, and then to enhance the optomechanical entanglement. We have also studied the universal physical coupling configuration for generation of optomechanical entanglement in the network-coupled four-mode optomechanical system. The results show that the generation of optomechanical entanglement depends on the breaking of the dark mode. To break the dark mode formed in this physical model, an efficient way is to introduce an auxiliary cavity mode and couple it with either one of the two mechanical modes. This scheme is experimentally accessible with a microwave optomechanical setup. Our results will not only pave a way toward the generation and manipulation of macroscopic optomechanical entanglement, but also initiate the study of noise-immune quantum resources.

\begin{acknowledgments}

J.-Q.L. was supported in part by National Natural Science Foundation of China (Grants No. 12175061, No. 11822501, No. 11774087, and No. 11935006), the Science and Technology Innovation Program of Hunan Province (Grants No. 2021RC4029 and No. 2020RC4047), and Hunan Science and Technology Plan Project (Grant No. 2017XK2018).

\end{acknowledgments}

\end{document}